\documentclass[conference]{IEEEtran}
\usepackage{amsmath,amssymb,amsfonts}
\usepackage{algorithm}
\usepackage{algorithmic}
\usepackage{graphicx}
\usepackage{textcomp}
\usepackage{xcolor}
\usepackage{balance}
\usepackage{subfigure,wrapfig}
\usepackage{comment}
\usepackage{subfigure, color, colortbl}
\newcommand*{\revise}{\textcolor{black}}
\newcommand*{\edit}{\textcolor{black}}

\def\BibTeX{{\rm B\kern-.05em{\sc i\kern-.025em b}\kern-.08em
    T\kern-.1667em\lower.7ex\hbox{E}\kern-.125emX}}
    
\usepackage[hyphens]{url}
\usepackage[hidelinks]{hyperref}
\usepackage[noadjust]{cite}
\hypersetup{breaklinks=true}

\newcommand\blfootnote[1]{%
  \begingroup
  \renewcommand\thefootnote{}\footnote{#1}%
  \addtocounter{footnote}{-1}%
  \endgroup
}

\newcommand{\redundantResidualEnergy}{superfluous residual energy }

\newtheorem{definition}{Definition}

\IEEEoverridecommandlockouts\IEEEpubid{\makebox[\columnwidth]{ 978-1-7281-6992-7/20/\$31.00 $\copyright$2020 IEEE \hfill}\hspace{\columnsep}\makebox[\columnwidth]{ }}

\begin{document}

\title{Long-Lived LoRa: Prolonging the Lifetime of a LoRa Network \\
}

\author{\IEEEauthorblockN{Sezana Fahmida\IEEEauthorrefmark{2}\IEEEauthorrefmark{4}, Venkata P Modekurthy\IEEEauthorrefmark{2}\IEEEauthorrefmark{4}, Mahbubur Rahman\IEEEauthorrefmark{3},
		Abusayeed Saifullah\IEEEauthorrefmark{4}, Marco Brocanelli\IEEEauthorrefmark{4}} 
	%\IEEEauthorblockA{Department of Computer Science, Wayne State University, Detroit, MI, USA}
	\IEEEauthorblockA{\IEEEauthorrefmark{4} Wayne State University and \IEEEauthorrefmark{3} CUNY Queens College
		%\IEEEauthorrefmark{3}Department of Computer Science, Missouri University of Science and Technology, Rolla, MO, USA\\
		%\IEEEauthorblockA{\IEEEauthorrefmark{2}Wayne State University and \IEEEauthorrefmark{3}Missouri University of Science \& Technology
		%\{modekurthy,saifullah\}@wayne.edu,
		%madrias@mst.edu
		%\vspace{-0.2in}
        }
        }

%\IEEEauthorblockN{Sezana Fahmida\IEEEauthorrefmark{2},
%	\IEEEauthorblockN{Venkata P Modekurthy\IEEEauthorrefmark{2},
%	\IEEEauthorblockN{Mahbubur Rahman\IEEEauthorrefmark{2},
%		Abusayeed Saifullah\IEEEauthorrefmark{2} and Marco Brocanelli\IEEEauthorrefmark{2}} 
%	\IEEEauthorblockA{\IEEEauthorrefmark{2}Department of Computer Science, Wayne State University, Detroit, MI, USA

%\author{\IEEEauthorblockN{Sezana Fahmida
%\IEEEauthorblockA{
%\textit{Wayne State University}\\
% \\
%}
%\and
%\and
%\and
%\and
%\and

\maketitle

\pagestyle{plain}

\begin{abstract}

%Today, LoRa (Long Range) is a leading low-power wide-area network (LPWAN) technology that can enable communication between numerous devices (sensor node)  and a gateway over long distances at low cost and low energy consumption, thus enabling many wide-area Internet-of-Things (IoT) applications. Typically, in these applications battery-powered sensor nodes are deployed to provide uninterrupted service for a long interval. Recently, energy-harvesting technologies  have emerged as an efficient alternative to traditional batteries.In this paper, we propose the first method to maximize the network lifetime in a generic setup that includes an energy-harvested LoRa network. For LoRa network, existing works have studied maximizing network lifetime in the context of finding the minimum energy consuming transmission parameter.

Prolonging the network lifetime is a major consideration in many Internet of Things applications. In this paper, we study maximizing the network lifetime of an energy-harvesting LoRa network. Such a network is characterized by heterogeneous recharging capabilities across the nodes that is not taken into account in existing work. We propose a link-layer protocol to achieve a long-lived LoRa network which dynamically enables the nodes with depleting batteries to exploit the superfluous energy of the neighboring nodes with affluent batteries by letting a depleting node offload its packets to  an affluent node. By exploiting the LoRa's capability of adjusting multiple transmission parameters, we enable low-cost offloading by depleting nodes instead of high-cost direct forwarding. Such offloading requires synchronization of wake-up times as well as transmission parameters between the two nodes which also need to be selected dynamically. The proposed protocol addresses these challenges and prolongs the lifetime of a LoRa network through three novel techniques. (1) We propose a lightweight medium access control protocol for peer-to-peer communication to enable packet offloading which circumvents the synchronization overhead between the two nodes. (2) We propose an intuitive heuristic method for effective parameter selections for different modes (conventional vs. offloading). (3) We analyze the energy overhead of offloading and, based on it, the protocol dynamically selects affluent and depleting nodes while ensuring that an affluent node is not overwhelmed by the depleting ones. Simulations in NS-3 as well as real experiments show that our protocol can increase the network lifetime up to $4$ times while maintaining the same throughput compared to traditional LoRa network.

%In this paper, we study the network lifespan in a generic setup that includes energy-harvested LoRa nodes connected to one or more gateways for data analysis. For LoRa network, existing works have studied maximizing network lifetime in the context of finding the parameters such as spreading factor, coding rate, and transmission power. However, finding the right parameters does not always maximize the network lifetime. In some cases, a node may require to transmit more packets than its battery may support. As transmissions to neighboring nodes consume less energy, we propose to offload some of the transmissions from these nodes to a neighboring node which has a higher amount of energy left in its battery. To enable transmissions between neighboring nodes, nodes usually require transmission parameter coordination and synchronization, which is undesirable for low-power devices. To circumvent this, we propose a MAC protocol and parameter assignment for enabling packet offloading. If and when such offload should occur to/from a node is a critical decision. We propose a greedy heuristic algorithm that selects the nodes and the offloading duration. Simulation and experiment results show that our approach outperforms traditional LoRaWAN.
\end{abstract}

%\begin{IEEEkeywords}
%component, formatting, style, styling, insert
%\end{IEEEkeywords}

%to do point out differences with existing work
%to do motivate the problem
%to do motivate the need to use machine learning approach 
\blfootnote{\IEEEauthorrefmark{2}Co-first-author}
%\footnote{Co-first-author}
%!TEX root = main.tex

\section{Introduction}\label{sec:intro}

LoRa (Long Range) is a leading low-power wide-area network (LPWAN) technology that enables low-power (milliwatts) wireless devices to transmit at low data rates (kbps) over long distances (kms) using narrowband (kHz) \cite{lora}.   It can be deployed for direct communication between numerous end devices (also called {\slshape sensor nodes}) and a gateway in many wide-area Internet of Things (IoT) applications including smart agriculture~\cite{agricultureIoT}, smart city~\cite{airIoT}, and environmental monitoring~\cite{wildfireIoT}. While these end devices are usually powered by traditional single-use batteries, energy-harvesting technologies that exploit the light \cite{solarIoT}, the vibrations \cite{vibration}, and/or the heat \cite{heatIoT} of the environment have emerged as an efficient alternative (e.g., in smart agriculture, environmental monitoring) for providing sustainable energy. The carbon footprint of such energy-harvesting technologies is many times lower than the other energy sources including single-use batteries 
 \cite{carbonfootprint, batterysecret}.  However, relying on energy-harvesting for recharging the sensor nodes can lead to heterogeneous state of charge for the batteries in the network. Thus, it is crucial to regulate the energy consumption of the nodes dynamically to ensure maximum network availability.

%%motivating the problem
%LoRa (Long Range) is a leading low-power wide-area network (LPWAN) technology that can enable direct communication between numerous devices and a gateway over long distances at low cost and low energy consumption~\cite{lora}. It can be employed in many wide-area Internet of Things (IoT) applications, including smart agriculture~\cite{agricultureIoT}, smart city~\cite{airIoT}, and environmental monitoring~\cite{wildfireIoT}. In these applications, the devices (e.g. sensor nodes) transmit data to a server for further processing. Usually, these devices are powered by traditional single-use batteries. However, these batteries require regular human intervention~\cite{8362652}. Recently energy-harvesting technologies which exploit the light \cite{solarIoT}, the vibrations \cite{vibration}, and/or the heat \cite{heatIoT} of the environment to provide sustainable energy to the nodes (e.g., in smart agriculture,   environmental monitoring) have emerged as an efficient alternative to traditional batteries. Besides, the carbon footprint of such energy-harvesting technologies is many times lower than the other energy sources including batteries 
% \cite{carbonfootprint, batterysecret}.  However, relying on energy-harvesting for recharging the sensor nodes can lead to heterogeneous levels of batteries in the network. Thus, it is crucial to regulate the energy consumption of the nodes dynamically to ensure maximum network availability. 

Prolonging the network lifetime is a major consideration in most IoT applications. We consider {\slshape network lifetime} as the interval starting at any given point in time until the first  node depletes its battery. In many IoT applications the lifetime directly implies the availability of the services as failure to collect sensor data from even one node may lead to unmet application requirements. In this paper, we propose the first method to maximize the network lifetime in a generic setup of an energy-harvesting LoRa network \cite{wu2018we,sherazi2018renewable}.

Recently there has been some research focusing on adopting a general energy minimization strategy for all nodes to ensure the energy-efficiency of LoRa networks \cite{piyare2018demand, gao2019towards}. However, following the same energy minimization strategy for all nodes may not always be beneficial for prolonging the lifetime in a network where nodes may have heterogeneous recharging capabilities. Insufficient energy generation in a recharge cycle may lead to an unexpected depletion of a node's battery even with the most energy-efficient transmission parameter allocation. Hence, in order to maximize the network lifetime it is required to have an approach that prevents such depletion. 
For example, a \textit{depleting} node that has not harvested enough energy can drain its battery rapidly if proper action is not taken. However, there may be \textit{affluent} nodes in the network that have \redundantResidualEnergy in their batteries that will remain unused in the current recharge cycle. We propose a link-layer protocol to maximize the network lifetime by dynamically enabling the depleting nodes to exploit the superfluous energy of the neighboring affluent nodes.

%Furthermore, existing works~\cite{gao2019towards} have provided a general energy minimization strategy for all nodes in the network, which may not always be beneficial for extending the lifetime in a network where nodes have recharging capabilities. 

%high level view of our approach
%For example, node $a$ (\textit{depleting node}) may not generate enough energy in the current cycle and may run out of energy, whereas node $b$ (\textit{affluent node}) may generate surplus energy and may not use it in the current cycle. 
%Following the same energy minimization strategy for all nodes as typically adopted in existing work may not be advantageous in terms of lifetime maximization. 

In our approach, the superfluous energy of an affluent node is exploited by letting a depleting node {\slshape offload} its workload to the former. Namely, if
the two nodes are within communication range of each other, a depleting node offloads its packets when needed to an affluent node which then transmits those to the gateway. In a LoRa network, transmission to a close-by node can be made at much less energy by
adjusting {\slshape multiple} transmission parameters compared to directly forwarding to the gateway. We exploit this feature of the LoRa technology and enable low-cost (in energy) offloading by depleting nodes instead of high-cost direct forwarding. Note that, in this strategy, the overall energy consumption at the affluent node and, possibly, the total energy needed for delivering a packet to the gateway may be higher than that needed for direct forwarding from the depleting node.
However, as counter-intuitive as it may seem, it is very effective in maximizing the lifetime of the network because it prevents the depleting node from rapidly consuming its battery. %On the other hand, to correctly realize this intuition, we need to dynamically select  depleting and affluent nodes and ensure that the latter are not overloaded. 
This idea forms the central thesis of this paper to achieve a long-lived LoRa network.

%As counter-intuitive as it may seem, we make the observation that, in some cases, \emph{a higher energy consumption to deliver packets from a node to the gateway can actually help increase the network lifetime}. By adjusting multiple transmission parameters, a depleting node without enough energy to transmit directly to the gateway can make a transmission to a close-by affluent node at much less power than transmitting directly to the gateway. Thus, a depleting node can \textit{offload} its packet to a near-by affluent node, which can make the transmission to the gateway on the depleting node's behalf. Note that, in this strategy, the overall energy consumption at the affluent node increases and also the total energy needed for delivering a packet to the gateway may be higher than that needed for directly transmitting to gateway from the depleting node. However, it is very effective in maximizing the lifetime of the network as it prevents the depleting node from rapidly consuming its battery as long as we properly select the depleting and affluent node dynamically and ensure that the latter are not overloaded. This idea forms the central thesis of this paper for maximizing the lifetime of a LoRa network.

Enabling packet offloading between nodes  in a LoRa network raises three main challenges. First, the energy-efficient offloading of packets requires synchronization of wake-up times between the nodes, which typically introduces additional energy overheads. Second, a successful offloading requires that the nodes be operating on the same transmission parameters. 
Thus, we need to decide energy-efficient transmission parameters to enable offloading. Third, we need to dynamically select depleting and affluent nodes without overwhelming the latter, while also preventing the rapid battery depletion of the former.

We address the above challenges by developing the proposed link-layer protocol which enables a long-lived LoRa network. It prolongs the lifetime of a LoRa network through three novel techniques. (1) We propose a lightweight Medium Access Control (MAC) protocol for peer-to-peer communication to enable packet offloading which circumvents the synchronization overhead between the two nodes, thereby ensuring low power consumption and reliable packet delivery during offloading. (2) We propose an intuitive heuristic method for parameter selections for different modes (conventional vs. offloading) to ensure that only the required energy for successful transmission is spent in each mode. (3) We analyze the energy overhead for an affluent node when a packet is offloaded to it. Based on this analysis, the protocol dynamically selects affluent and depleting nodes during network operation as well as the duration in such modes so that an affluent node is not overwhelmed by the depleting ones. 
We have evaluated the proposed protocol through extensive simulations in NS-3 and physical experiments. The results show that it can increase the network lifetime up to $4$ times while maintaining the same throughput compared to traditional LoRa network.

%(2) We realize the lifetime maximization by proposing three different modes of operation, namely \textit{lading}, \textit{offloading}, and \textit{conventional}. 

In the rest of the paper, Section~\ref{sec:background_model} describes background and model. Section~\ref{sec:relatedwork} describes  related work. Section~\ref{sec:longlivedlora} describes the protocol to prolong the network lifetime. Section~\ref{sec:evaulation} provides the evaluation results. Section~\ref{sec:conclusion} concludes the paper.

%!TEX root = main.tex

\section{Background and System Model} \label{sec:background_model}
%An LPWAN enables low data rate and low-power communication directly between several nodes and one or more gateways over long distances (e.g., several miles). 

%We consider an LPWAN based on the LoRa technology. 

%Section \ref{sec:lora} presents an overview of the LoRa modulation and LoRa wide area network. Section \ref{sec:model} presents the system model under consideration for this paper.

\subsection{An Overview of LoRa and LoRaWAN}
\label{sec:lora}
LoRa is a physical layer technology for LPWAN with a communication range of 3-10 miles~\cite{lora} (depending on the environment). Modulation in LoRa is derived from the {\slshape chirp spread spectrum (CSS)}. In CSS, one bit of information or chirp is a signal continuously varying in frequency, often represented as multiple chips. Spreading the frequency of the signal enables the successful reception of a packet at long ranges and low/negative signal-to-noise ratios. The amount of frequency spreading applied to the signal in a chirp is called the \textit{spreading factor} of transmission. LoRa adopts seven orthogonal spreading factors in the range $[6, 12]$ for simultaneous reception on the same channel. Each spreading factor is a binary log-scale representation of the number of chips present in one symbol, i.e., spreading factor $6$ represents $64$ chips are present in one symbol. In CSS, apart from spreading factors, nodes can be configured to use different bandwidth settings, channel, coding rates, transmission powers.

In North America, LoRa supports sixty-four channels with a maximum bandwidth of $125$kHz and eight channels with a maximum bandwidth of  $500$kHz for uplink communication between a node and a gateway. Additionally, it supports eight downlink channels with a maximum bandwidth of $500$kHz. 
%LoRa supports communication on different bandwidths with the lowest bandwidth of 7.8kHz. 
A packet in LoRa communication supports four levels of forward error detection and correction mechanisms called {\slshape coding rate}. Coding rate ranges from $\frac{4}{5}$ to $\frac{4}{8}$. Lower coding rates enable better error detection and correction mechanism; however, they increase the packet size and energy consumption.

%\begin{figure}
%   % \vspace{-2mm}
%    \centering
%    \includegraphics[width=0.3\textwidth]{Fig/network_model.eps}
%        \vspace{-3mm}
%    \caption{LoRa Network Architecture}
%    \label{fig:LoRa_Architecture}
%    \vspace{-5mm}
%\end{figure}

\begin{wrapfigure}{l}{4.7cm}
   \vspace{-3mm}
    \centering
    \includegraphics[width=0.27\textwidth]{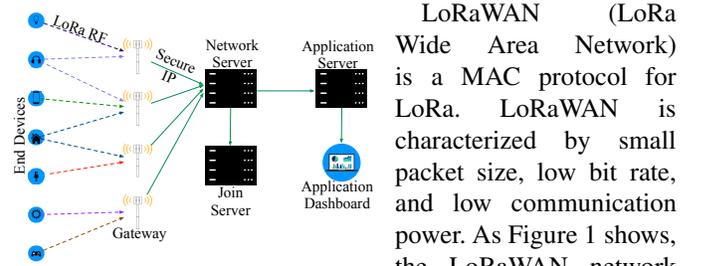}
        \vspace{-6mm}
    \caption{LoRa Network Architecture}
    \label{fig:LoRa_Architecture}
    \vspace{-5mm}
\end{wrapfigure}
LoRaWAN (LoRa Wide Area Network) is a MAC protocol for LoRa. LoRaWAN is characterized by small packet size, low bit rate, and low communication power. As Figure \ref{fig:LoRa_Architecture} shows, the LoRaWAN network consists of nodes/end-devices, gateway, network server, application server, and a join server. A {\slshape gateway} acts as a relay between wireless communication (with the nodes) and wired communication (with the network server). The {\slshape network server} manages the network parameters, security, and application requirements. The {\slshape application server} interprets sensor and application data from the sensors. Nodes in the network initiate a join request upon deployment, which is handled by the {\slshape join server}. 

LoRaWAN supports three classes of operation. For the uplink communication, LoRaWAN adopts pure ALOHA for transmission in all three classes, where a node transmits whenever it has a packet. In {\slshape class A}, every transmission precedes two receive time windows for successfully receiving an acknowledgment (ACK). In {\slshape class B}, the gateway periodically transmits a beacon for time synchronization. Between two beacons, a node periodically wakes up to receive any packets sent by the gateway. In {\slshape class C}, nodes are continuously listening to a packet transmission from the gateway. %Class C mode has the lowest latency for downlink communication but the highest energy consumption, while class A has the highest latency for downlink communication but the lowest energy consumption. 

\subsection{System Model}
\label{sec:model}
The nodes in the network are powered by batteries recharged through energy harvesting devices (e.g., solar panels). The nodes collect and transmit sensor information to the gateway. They typically use class A mode for transmitting their packets to the gateway and sleep for the rest of the time to conserve energy. We denote the transmission parameters such as spreading factor, channel, coding rate, and bandwidth of a node $v$  by $SF_v$, $C_v$, $CR_v$, and $BW_v$, respectively. 

The \textit{line-powered} gateways listen for packets, acknowledge them, and forward them to the network server and application server. In addition to providing security, the network server stores the location and current battery level of each network node. Because node localization is out of scope for this paper, we assume it is performed at network deployment time through manual configuration or some existing localization techniques for low-power networks \cite{wsnlocalizationsurvey}. The battery level of each network node is embedded in the existing uplink packet to the network server for timely state of charge updates. %If a node does not have any data to send for a long time, it can use a heartbeat message to keep the node informed about its energy level.

Typically, the nodes periodically sense the state of the system. However, in some cases, the sensed information may not change over time; hence, some node% does not need to send the information to the gateway. Thus, the node
s may locally determine if the sensed data needs reporting,   %For example, in a smart farming application, if the soil moisture does not change from the last reading, the nodes need not communicate the information to the gateway, thereby saving energy. 
which leads to aperiodic transmissions with a known minimum inter-arrival time between packets.  When a node does not have any packets to transmit for long intervals, it communicates a heartbeat message with the gateway to maintain connectivity. The period of the heartbeat message is set by the network manager.

%Typically, nodes with mutually exclusive sensing capabilities may exist within a small neighborhood. For example, in a smart city application, network deployment may include two nodes $u$ and $v$ within a block of each other, where $u$ measures temperature, humidity, and wind speed, and $v$ measures carbon dioxide levels, carbon monoxide, and other harmful gasses. Typically, nodes with different sensing capabilities have different communication requirements. 

%When a node v transmits packets related to a sensor data it consumes energy as follows:
\textbf{Transmission Energy Model.} The energy consumed for packet transmissions from a node $v$ can be modeled as follows:
\begin{equation}
\label{eq:energy_transmission}
E_{\text{tx}} (v) =  P_{\text{tx}} (PL_v) \times N_{\text{symbols}} (v) \times T_{\text{symbol}} (v) 
\end{equation}
%Energy consumed by a node $v$ during a packet transmission in LoRa is given by the Equation (\ref{eq:energy_transmission}).  
where $P_{\text{tx}} (PL_v)$ represents the power consumed by the LoRa chip to transmit a packet at power level $PL_v$, $T_{\text{symbol}} (v)$ represents the time required to transmit one symbol with spreading factor $SF_v$ and bandwidth $BW_v$, and $N_{\text{symbols}}(v)$ denotes the number of symbols in a packet. 
The number of symbols in a packet from $v$ including a header and cyclic redundancy check (CRC) code is modeled as follows: 
\begin{equation*}
%\label{eq:number_of_symbols}
\begin{split}
N_{\text{symbols}} (v) =& preamble_v + 4.25 + 8 + \\& \max \bigg( \bigg\lceil  \frac{8 \, payload_v - 4\,SF_v + 24}{SF_v - 2\, DE_v}  \bigg\rceil \frac{1}{CR_v}, 0 \bigg)
\end{split}
\end{equation*}
where  $preamble_v $ represents the length of the preamble and $payload_v$ represents the length of the payload. Equation (\ref{eq:symbol_duration}) models $T_{\text{symbol}} (v)$ and Equation (\ref{eq:de}) describes the low data rate optimization mode $DE_v$:
\vspace{-2mm}
\begin{equation}
\label{eq:symbol_duration}
T_{\text{symbol}} (v) = \frac{2^{SF_v}}{BW_v}
\end{equation}
\begin{equation}
\label{eq:de}
DE_v = \begin{cases}
        1, & \text{if low data rate optimization is enabled} \\
        0, & \text{otherwise}
        \end{cases}
\end{equation}

During a receive window of length $T_{\text{rx}} (v)$ seconds, node $v$ consumes energy as follows:
\vspace{-2mm}
\begin{equation}
\label{eq:energy_reception}
E_{\text{rx}} (v) =  P_{\text{rx}} (DE_v) \times  T_{\text{rx}} (v)
\end{equation}
where $P_{\text{rx}} (DE_v)$ is the power consumed by the LoRa chip.
%This paper assumes that the applications use drones with an array of batteries and directional wireless chargers recharge the batteries of nodes periodically. Such a drone-based charging of batteries at nodes eliminates any need for a line power. Drones use energy from green sources, such as solar panels and wind farms, to charge its array of batteries. In cases where sufficient green energy sources do not generate adequate electricity, applications may use line power to charge the batteries on the drone. 

%%%Comment - draw a figure of how the charge rate changes

%%%Comment - replace T^l with a different notation
%\begin{figure}[th]
%    \centering
%    \vspace{-2mm}
%    \includegraphics[width=0.3\textwidth]{Fig/solarpower.eps}
%    \caption{Hourly average solar power generation}
%    \label{fig:solarpowergeneration}
%    \vspace{-2mm}
%\end{figure}

\textbf{Recharge Energy Model.} Each network node is equipped with an independent green energy source that recharges the node's local battery. Typically, the \edit{instantaneous power} generated by any green source varies over time but repeats over a fixed time interval of length $\zeta$ called a {\slshape recharge cycle}. A recent study has shown that the average power generated for one hour interval by a solar panel repeats every 24 hours \cite{jahid2019hybrid}. Thus, %from the past data, 
the network server can estimate the \edit{energy} generated at node $v$ during a recharge cycle as \edit{the integral of the estimated instantaneous power ($P_{\text{recharge}} (v)$) over the recharge cycle, as shown in Equation (\ref{eq:charge_accumulated})}. 
%\begin{equation} \label{eq:charge_accumulated}
%B_v^* = P_{\text{recharge}} (v) \times \zeta
%\end{equation} 
\vspace{-2mm}
\begin{equation} \vspace{-2mm}
\label{eq:charge_accumulated}
B_v^* =  \int_{0}^{\zeta} P_{\text{recharge}} (v) dt
\end{equation} 
%Node $v$'s energy consumption budget for the interval of length $\zeta$ is the same as $B_v^*$, the estimated average energy generated during the same period. 
%The network server can synchronize the start and end of the recharge cycle at all nodes. 
When the network is deployed, the batteries are usually fully charged. Typically, the energy harvesting system is conservatively designed to be sized (1) proportionally to the battery size of each node, i.e., the energy stored in a fully-charged battery is no less than $B_v^*$, and (2) so that the average energy consumed during a recharge cycle is no greater than the average recharged energy during the same period. %In this conservative scenario, a depleted battery may need more than one recharge cycle of length $\sigma$ to completely recharge. In addition, within an interval of length \sigma, neighboring nodes can have different average energy generation due to environmental effects like shadows impact the energy generation. 
\edit{Thus, we define {\slshape battery budget} of a node $v$ during $\zeta$ time units as $B_v^*$, an estimation of the energy generated during that interval. }\revise{The network server synchronizes the start and end of the recharge cycle to ensure all nodes reset the battery budget at the same time and perpetuate network communication.}
In this conservative scenario, within an interval of length $\zeta$, neighboring nodes can have different \edit{battery budgets} %average energy generation 
due to environmental effects such as shadows that impact the energy generation. Similarly, nodes can have different energy consumption due to the heterogeneity of hardware and/or communication patterns.

%When the network is deployed, the battery of each node is usually fully charged and hence the battery level at node $v$ at the deployment time is no less than  $B_v^*$. %During the interval of length $\zeta$ time units, node $v$ consumes at most $B_v^*$ energy but generates at least $B_v^*$ energy. 
%Since a node consumes energy less than or equal to the energy generation within a recharge cycle, the battery is restored to a similar level to the start of the previous cycle. Thus assuring the continuous operation in the subsequent recharge cycles.   %This assumption enables the use of different instantaneous energy consumption and generation rates of the battery at a node. 

%Within an interval of length $\zeta$, neighboring nodes can have different average energy generation due to environmental effects like shadows impact the energy generation. Similarly, nodes can have different energy consumption due to the heterogeneity of communication patterns.

\textbf{Network Lifetime.} We calculate the network lifetime as the time interval from the start of the recharge cycle until the first node consumes its battery budget. Although there are other definitions of network lifetime, we choose this definition of network lifetime as it hampers the data collection from at least one node in the present or the future intervals. Given the above considerations, we regulate the nodes' energy consumption in each interval according to the estimated $B_v^*$ energy recharged within each interval $\zeta$. This strategy ensures (1) there is no downtime in the current interval and (2) the batteries' energy level at the end of the current interval is higher than a certain  level necessary to ensure continuous operations for the next.

%Given the network model and lifetime definition, the objective of this paper is to maximize the network lifetime within each recharge cycle of length $\zeta$. Specifically, for the rest of the paper, we focus on maximizing the network lifetime in the interval [0, $\zeta$] since extending the lifetime beyond $\zeta$ in the first interval ensures the same in subsequent intervals.
%!TEX root = main.tex

\section{Related Work}
\label{sec:relatedwork}
%As also discussed in a recent survey~\cite{yetgin2017survey}, there has been a plethora of work on lifetime maximization for wireless sensor networks. However, these approaches consider multi-hop mesh networks whereas in a LoRa network each node is connected directly to the gateway. Furthermore, LoRa has multiple configurable transmission parameters to control the energy consumption of the nodes in the network, thus it needs a different approach to achieve longer lifetime.
%There has been a plethora of work on lifetime maximization for wireless sensor networks ~\cite{yetgin2017survey}. However, these works are not applicable for maximizing the lifetime of the network for LoRa since they do not leverage on the simultaneous reception of packets from multiple nodes, available in LoRa. Leveraging on the simultaneous reception significantly reduces the interference from other nodes, which can maximize the lifetime of the network. Furthermore, LoRa has multiple configurable transmission parameters that impact the energy consumption of the nodes in the network.

\revise{Due to their ability to overcome the coverage and scalability limitations in traditional short-range wireless sensor network \cite{capnet}, LPWANs have received considerable attention recently \cite{raza, snow2, lpwanicdcn18, iotdi2018, Charm, choir, iotdi2019}.  While many LPWANs have been designed (e.g, LoRa \cite{lora}, SigFox \cite{sigfox}, IQRF \cite{iqrf}, RPMA \cite{rpma}, DASH7 \cite{dash7}, Telensa \cite{telensa}, NB-IoT~\cite{nbiot},   SNOW \cite{ton_snow, snow}, and LTE Cat M1 \cite{cat, LTE_advancedpro}), LoRa has gained popularity due to its wide availability. It has been the focus of much research in both academia and industry \cite{overview1}. A comprehensive review of these works can be found in \cite{lora_survey_2020}. Most of these existing works have focused on  enhancing     throughput \cite{wangmlora,xia2019ftrack}, latency \cite{wu2019distributed,leonardi2019rt},  reliability \cite{LoRaReliability, Charm, marcelis2017dare, pham2019investigating, LoRaInterference}, or scalability and coverage \cite{bor2016lora, slabicki2018adaptive, choir, LoRaScalability1, van2017scalability} of a LoRa network.}

\revise{Energy efficiency of a LoRa network was studied in \cite{piyare2018demand} using additional hardware and in \cite{gao2019towards, LoRaEnergyevaluation} by minimizing energy consumption at all nodes through transmission parameters selection \cite{gao2019towards, LoRaEnergyevaluation}. However, these approaches are not applicable to maximize the network lifetime of an energy harvesting LoRa network. Insufficient energy generation in a recharge cycle may lead to an unexpected depletion of a node’s battery even with the most energy-efficient transmission parameters. On the contrary, our proposed approach minimizes such depletion, thereby maximizing the network lifetime.}

\section{Proposed Protocol for Long-Lived LoRa} 
\label{sec:longlivedlora}

%\subsection{An Overview of the protocol}
%\label{sec:proposedmethod}

For nodes with battery recharging capabilities, minimizing energy consumption is not always beneficial for maximizing the lifetime of the network.  For example, if a node is fully charged at time $0$ and is expected to generate $B^*$ units of energy by time $\zeta$,  then consuming any amount of energy in the range [0, $B^*$] will still result in a full battery at time $\zeta$. 
%Thus, for nodes with battery recharging capabilities, ensuring energy consumption is less than or equal to the expected energy generation (or battery budget) maximizes the network lifetime.
Thus, the network lifetime is not negatively affected as long as it is ensured that a node's energy consumption is no greater than its expected energy generation (or battery budget). 

%This paper relies on this intuition and proposes a packet offloading mechanism. In the packet offloading mechanism, nodes with depleting battery budget (i.e., nodes that do not have sufficient battery budget to transmit all packets to the base station) forward several packets to a neighboring node with an affluent battery budget. The neighboring node forwards these packets to the base station on behalf of the depleting node. Typically, packet transmission to the neighbors consumes less energy than that to the base station.  Thus, packet offloading conserves energy at the depleting nodes and thereby increases the network's lifetime.

We rely on the above observation to maximize the network lifetime by enabling the nodes to exploit \redundantResidualEnergy of neighboring nodes. This is done through a packet offloading mechanism where a node whose battery budget is insufficient to deliver all subsequent packets directly to the gateway dynamically forwards a subset of its future packets to a neighboring node with an affluent battery budget. The neighboring node forwards these packets to the gateway on behalf of the depleting node. %In a LoRa network, 
\revise{Since we select neighboring nodes that are significantly closer to the node than the gateway, transmission to a neighboring} node can be made at much less power by adjusting multiple transmission parameters than transmitting directly to the gateway. Note that, in this strategy, the overall energy consumption at an affluent node increases and the total energy needed for delivering a packet to the gateway may be higher than that required for direct transmission to the gateway from a depleting node. However, it conserves energy at the depleting node. Thus, it helps prolong the network lifetime as long as we select the affluent and depleting nodes dynamically and ensure that the latter are not overloaded. 

\begin{figure}[th] %{r}{4.1cm}
    \centering%\hfill
     \vspace{-3mm}
      \subfigure[Network lifetime w/o offloading]{
        \label{fig:without_offloading}
        \includegraphics[width=.215\textwidth]{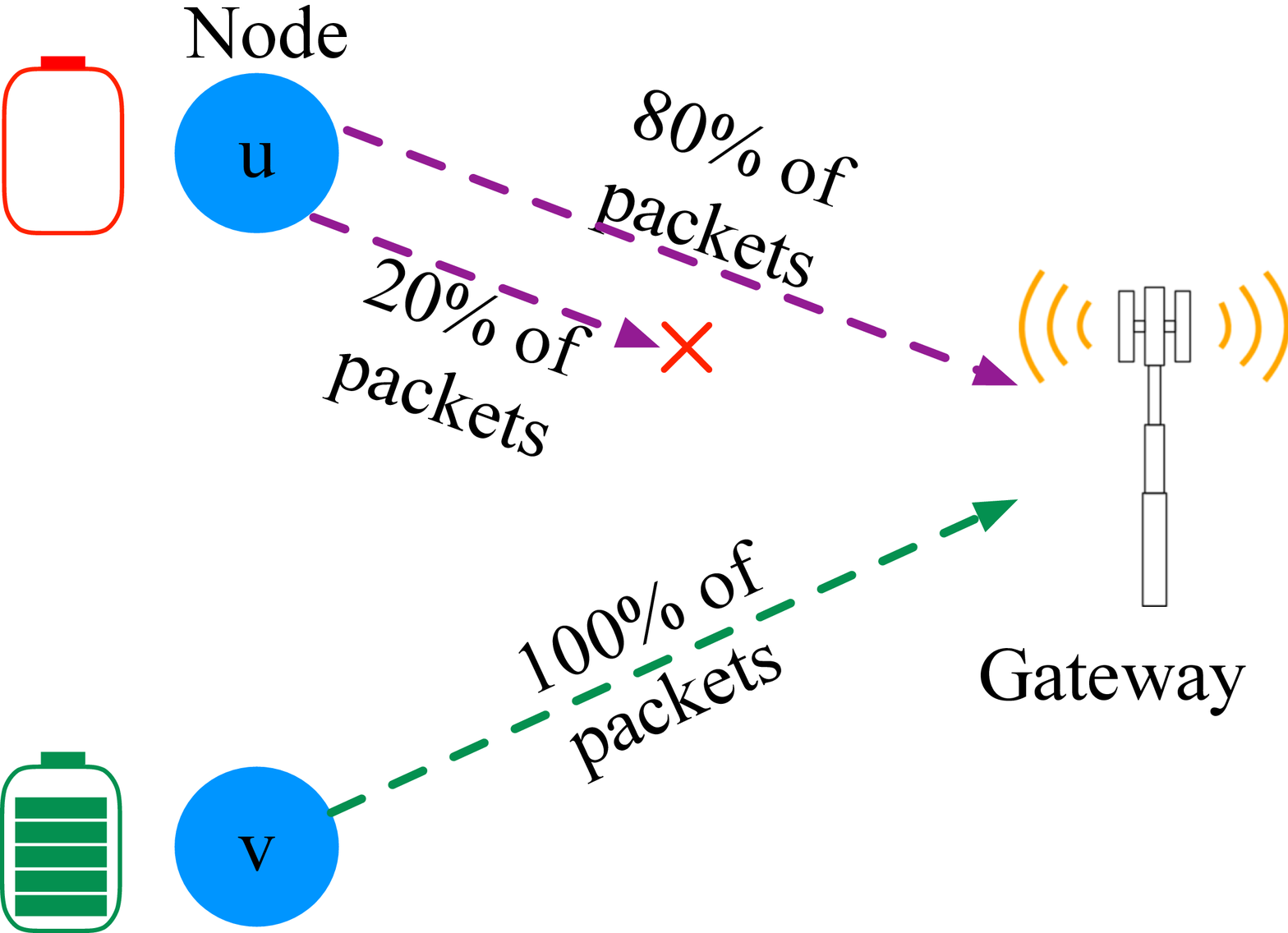}
      } \quad
      \subfigure[Network lifetime with offloading]{
      \label{fig:with_offloading}
    \includegraphics[width=0.215\textwidth]{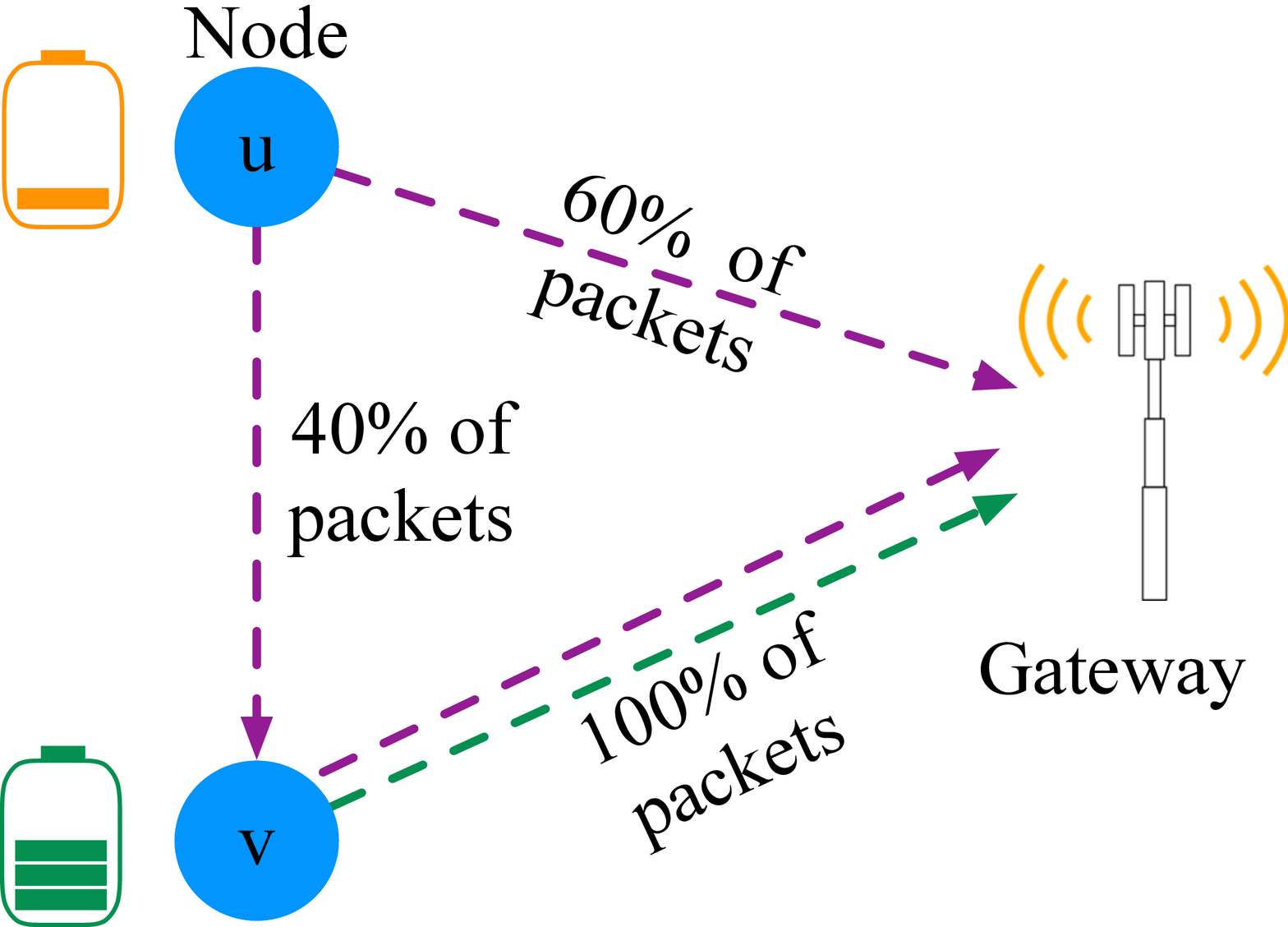}
      }
      \vspace{-0.12in}
    \caption{An example of maximizing network lifetime through packet offloading}%\vspace{-.15in}
    \label{fig:rl}
    \vspace{-2mm}
 \end{figure}

Fig. \ref{fig:rl} shows an example of the proposed offloading mechanism. In this example, two nodes, $u$ and $v$, transmit packets to the base station. Packet transmission from node $v$ consumes $10\%$ of its battery, while that from node $u$ consumes $100\%$ of its battery for transmitting only $80\%$ of the packets. In the traditional setting, $u$ runs out of charge before the next charging cycle. Assuming $u$ needs $50\%$ energy to transmit a packet to $v$ compared to the base station, offloading $40\%$ of $u$'s transmission to $v$ can ensure $u$'s (and thereby the network) lifetime is longer than the charging window.

%We formally define an affluent node and a depleting node in Definition (\ref{def:affluent}) and (\ref{def:depleting}) respectively.

\begin{definition}[\textbf{Affluent Node}]
  \label{def:affluent}
A node $v$ with an energy budget $B_v^*$ and an estimated energy consumption $B_v (t, \zeta)$ during the interval [$t$, $\zeta$], where $t$ is the current time and $\zeta$ is the start of next interval, is said to be an affluent node at time $t$ if and only if
 \vspace{-2mm}
\begin{equation}
\label{eq:affluent_node}
B_v (t, \zeta) < B_v^*.
\end{equation}

\end{definition}
\vspace{-1mm}
\begin{definition}[\textbf{Depleting Node}]
\label{def:depleting}
A node $v$ with an energy budget $B_v^*$, residual energy $\beta_v$ that needs to be saved for future, and estimated energy consumption $B_v (t, \zeta)$ during the interval [$t$, $\zeta$], where $t$ is the current time and $\zeta$ is the start of next interval, is a depleting node at time $t$ if and only if
 \vspace{-2mm}
 \begin{equation}
\label{eq:depleting_node}
B_v (t, \zeta) > B_v^* - \beta_v
\end{equation}
\end{definition}
\vspace{-1mm}
\noindent %In Fig. \ref{fig:rl}, node $u$ is a depleting node and node $v$ is an affluent node.
\revise{The value of $\beta_v$ depends on the estimated energy generation in the next recharge cycle of node $v$ and can be} estimated by the network manager during network deployment.
At time $t$, the estimate of energy consumption, battery budget for that interval, and Equations (\ref{eq:affluent_node}) and (\ref{eq:depleting_node}) are used to designate a node as affluent or depleting. 

\textit{Packet offloading}  from a depleting node $u$ to an affluent node $v$  in our approach means transferring the responsibilities of transmitting a set of packets to the gateway from $u$ to $v$. \edit{To enable packet offloading, we consider} three modes of operation for a node: \textit{offloading}, \textit{lading}, and \textit{conventional}. In the conventional mode, a node transmits packets directly to the gateway. In the lading mode, a node receives packets from a depleting node and forward them to the gateway. The offloading mode enables packet offloading to an affluent node. %At a time $t$, a node can be operating in any one of the three modes of operation. 

%In Fig. \ref{fig:without_offloading}, both nodes are operating in conventional mode. However, in Fig. \ref{fig:with_offloading}, depleting node $u$ operates $60\%$ of the time in conventional mode and $40\%$ of the time in offloading mode. Affluent node $v$ operates $60\%$ of the time in conventional mode and $40\%$ of the time in lading mode.

%\noindent \textbf{Impact of Packet Offloading}. 

Packet offloading increases the energy consumption of an affluent node but does not affect its lifetime when its total energy consumption is no greater than its battery budget. To ensure that the total energy consumption of an affluent node remains within its battery budget, we propose to find (i) the pairing between affluent nodes and depleting nodes that communicate with each other and (ii) an estimate of the operation time in lading and offloading mode for the affluent and depleting node, respectively. Since the solution to an optimal pairing of affluent-depleting nodes and operation time in lading and offloading mode is unknown, we propose a heuristic solution. 

The proposed heuristic for finding the affluent-depleting node pair that communicate with each other depends on the node's energy consumption in the lading mode. Energy consumption in the lading mode depends on the link-layer protocol that enables packet offloading. In the following sections, we describe the proposed link-layer protocol that enables packet offloading between the depleting and affluent nodes, transmission parameter selection, energy overhead in the lading mode, and selection of affluent and depleting node pairs that communicate with each other.

\subsection{Enabling Peer-to-Peer Communication for Offloading}
\label{sec:communication}
%Peer-to-peer communication requires synchronization of wake-up time and transmission parameters between the peers. Section \ref{sec:mac_lading} describes an approach to ensure lading nodes are in listen mode when offloading nodes are ready to transmit. Section  \ref{sec:parameter_selection} describes the proposed method to synchronize the transmission parameters between the lading node and offloading node.

\revise{Enabling communication between an affluent node and a depleting node requires synchronization between their sleep and wake-up times. Such a synchronization can incur huge energy overheads for both affluent and depleting nodes. To overcome this limitation, we adopt the low-power listening technique used in low-power network. In low-power listening, a node performs clear channel assessment (CCA) for a fixed duration to determine the channel activity. If the node detects channel activity, it wakes up and listens to the packet; otherwise, it sleeps. We propose to adopt a similar approach for the lading mode of LoRa. However, adopting low-power listening for LoRa has the following challenges: (1) using CCA to monitor the start of a packet can lead to several false-negatives since nodes can decode packets below the noise level, and (2) due to long communication range, a node can be in communication range with many other nodes, which can increase the number of false wake-ups in lading mode. We address these limitations by adopting channel activity detection (CAD) for low-power listening. The proposed peer-to-peer communication between an affluent and a depleting node is described in the following sections. First, we describe the communication in the lading mode and then in the offloading mode.}

%Abu's paragrpah
%We describe peer-to-peer communication between an affluent node and a depleting one in this section. First we describe \edit{the communication in} the lading mode and then in the offloading mode. 

% MAC protocol for LoRa. 
%Section \ref{sec:mac_transition} describes the protocol to initiate lading and offloading mode at nodes. %Finally, Section \ref{sec:lading_energy} presents the energy overhead for an affluent node in the lading mode. 

\subsubsection{Lading Mode for LoRa}
\label{sec:mac_lading}

\edit{In LoRa's lading mode,} the affluent node periodically wakes up and performs CAD.
%%%%%%%%comment the next line when commenting the new paragraph about false wake-ups <--------
\edit{Note that most of the existing and commercially available off-the-shelf LoRa communication chips such as SX1276, SX1277, SX1278, and SX1279  implement and support CAD \cite{sx1276_datasheet}.} 
During CAD, the affluent node receives signals for two symbols and probes for correlation between the received signal and a known preamble to identify any ongoing packet transmissions. \edit{If the affluent node detects a packet transmission, it continues listening to the packet transmission; otherwise, it sleeps. Because a node probes for correlation with a known symbol, CAD overcomes the false-negative detection of packets.} After receiving a packet, the affluent node forwards the packet to the gateway. \edit{Since there is a minimum inter-arrival time between two packets at a node, the affluent node can transmit to the gateway in the interval between two offloading packets without interfering the depleting node's communications. The affluent node can also use this interval to transmit its own packets.}

\begin{figure}
    \centering
    \includegraphics[width=0.32\textwidth]{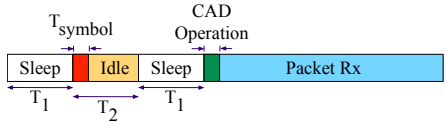}
    \vspace{-0.15in}
    \caption{Timing diagram in Lading Mode.}
    \vspace{-0.25in}
    \label{fig:lading_mode}
\end{figure}

Fig. \ref{fig:lading_mode} shows the timing diagram of a node operating in lading mode. An affluent node wakes up once every $T_\text{CAD} = T_1 + T_2$ seconds, where $T_1$ and $T_2$ are timer interrupts. % with the smallest value of $64 \mu s$. 
 The node sleeps for $T_1$ seconds. Within the $T_2$ seconds, the node performs CAD for $T_{\text{symbol}}$ seconds and \edit{remains in low-power idle mode} for the rest if a packet is not detected (for example, after the first CAD operation in Fig. \ref{fig:lading_mode}). If the node detects a packet during the CAD, for example, after the second CAD operation in Fig. \ref{fig:lading_mode}, it wakes up and listens to the packet.
%Most of the existing and commercially available off-the-shelf LoRa communication chips such as SX1276, SX1277, SX1278, and SX1279  implement and support CAD \cite{sx1276_datasheet}. Specifically, when set to 111, 0-2 bits of the RegOpMode register enable CAD on a LoRa SX1276 chip. If a CAD signal is detected, the SX1276 chip sets the 0th bit of the RegIrqFlag register to 1; otherwise, it resets the bit to 0. The node sets the 0-2 bits of the RegOpMode register to 110/101 to listen for packet transmissions upon verifying a packet reception. If CAD does not detect a  transmission, the node sets the 0-2 bits of  RegOpMode register to 000 to put the node to sleep. 
\revise{On the other hand, in LoRa communication the preamble for all packets is the same. Thus, during CAD an affluent node may not distinguish between transmissions to the affluent node and gateway, which would result in a significantly high number of false wake-ups. To solve this problem, we use reverse I-Q signals for packet transmissions to lading node and regular I-Q signals for packet transmissions to the gateway. During CAD, an affluent node can only listen to regular I-Q or reverse I-Q but not both, thus minimizing the number of false wake-ups.}

\subsubsection{Offloading Mode for LoRa}
\label{sec:mac_offloading}
The objective of the offloading mode is to minimize the energy consumption at a node, thereby increasing the network lifetime.  To meet this requirement, we propose to adopt a lightweight MAC protocol similar to LoRaWAN class A to send a packet. A node transmits a packet whenever it has data to send. If the affluent node acknowledges the packet reception in the next two time windows, it goes back to sleep until it has data to send. Otherwise, it retransmits the packet after a random back-off.

For a depleting node's packet to be successfully received by an affluent node through CAD, the duration of the preamble should be a minimum of $T_1 + 2T_2$. Since each symbol takes $\frac{2^{SF}}{BW}$, \edit{the length of the preamble is at  least $ \frac{(T_1 + 2T_2)BW}{2^{SF}}$. When $T_1 = T_2 = 4.1ms$ (the smallest feasible value possible for $T_1$ and $T_2$), the preamble should be at least $13$ symbols long for a spreading factor of $7$ and bandwidth of $125$kHz.}

%For a depleting node's packet to be successfully received by an affluent node through CAD, the duration of the preamble should be a minimum of $T_1 + 2T_2$. Since each symbols takes $\frac{2^{SF}}{BW}$, the length of the preamble is as follows:
%
%\begin{math}
%\label{eq:preamble_length}
%    \text{Preamble Length} \geq \frac{(T_1 + 2T_2)BW}{2^{SF}}
%\end{math}
%When $T_1 = T_2 = 4.1ms$ (the smallest feasible value possible for $T_1$ and $T_2$), the length of the preamble should be $13$ for spreading factor $7$ and bandwidth $125$kHz.

%In the offloading mode, a depleting node transmits a packet using the reverse I-Q signals, while in LoRaWAN, nodes use regular I-Q signal to transmit a packet to the gateway. CAD can detect packets either on regular I-Q signal or on inverse I-Q signal, but not on both.  Thus, an affluent node in the lading mode listening for inverse I-Q signals cannot listen to nodes in the conventional mode transmitting packets to the gateway. Packet transmissions using the reverse I-Q signals reduces the number of false wake-ups for the affluent node in the lading mode. A node can send a packet using a reverse I-Q signal by setting the $0$th bit of the RegInvertIQ register to $1$ \cite{sx1276_datasheet}. 

\subsection{Transmission Parameter Selection}
\label{sec:parameter_selection}
\edit{Packet offloading between nodes requires both the affluent node and the depleting node to operate on the same channel, spreading factor, bandwidth, and coding rate.} Due to the complex nature of this problem, we leave the optimal problem formulation and analyzing its complexity class as an open problem. In our approach, we develop a heuristic for transmission parameter selection in the conventional mode and offloading-lading mode to enable communication from node to gateway and 
from depleting node to affluent node, respectively.

%\begin{wrapfigure}{l}{7mm}
%    \centering
%    \includegraphics[width=0.15\textwidth]{Fig/channel_assignment.eps}
%    \caption{Example of Channel Assignment in the proposed approach, where each color represents a different transmission channel.}
%    \label{fig:my_label}
%\end{wrapfigure}

\begin{wrapfigure}{l}{3.8cm}
    \centering
    % \vspace{-2mm}
    \includegraphics[width=0.15\textwidth]{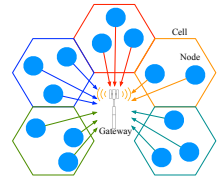}
     \vspace{-2mm}
    \caption{A channel assignment example (each color represents a different transmission channel).}
    \label{fig:my_label}
  \vspace{-1mm}
\end{wrapfigure} 
\noindent{\textbf{Channel Assignment. }} The objective of channel allocation is to ensure potential neighboring affluent and depleting nodes are assigned the same channel. The network server segregates the networks into cells and assigns the same channel to all the nodes within a cell. Furthermore, adjacent cells are assigned different channels to minimize the interference from depleting nodes in neighboring cells. Since neighboring nodes can have different communication patterns and battery budgets, the nodes within the same cell can have both depleting and affluent nodes. 
%Thus, this approach enables communication between affluent and depleting nodes. 
Fig.~\ref{fig:my_label} shows an example of our channel allocation. To generate cells, the network server uses location-based segregation of nodes. Initially, the network server creates cells equal to the number of channels. \edit{At the end of each recharge cycle, it can further divide} one cell into smaller cells if the number of affluent and depleting nodes within one cell is above a certain threshold. The network manager can set this threshold based on required performance improvement in the network lifetime. The same channel assignment is used in all operation modes. % of the nodes.

\noindent{\textbf{Transmission power. }}
In conventional mode, the nodes use transmission power based on ADR (adaptive data rate) adopted in LoRaWAN. In ADR, the nodes initially use the maximum transmission power to transmit a packet~\cite{LoRaWAN_spec_v1.1}. After every 20 packets received by the gateway, the nodes decrease the transmission power by 3dBm to compute the minimum transmission power that results in a successful reception. We use a similar approach in the lading-offloading mode, where an affluent node starts with a transmission power of 
\begin{math}
\label{eq:offloading_mode_initial_transmission_power}
PL_i = P_s + L_0 + 10n\log\frac{d}{d_0}, 
\end{math} where $P_s$ is the receiver sensitivity, $L_0 = 7.7$dBm, the path loss at reference distance $d_0 = 1$m, and $n = 3.76$ is the path loss exponent.
An affluent node decreases the transmission power if the signal to noise ratio (SNR) is significantly higher than the receiver sensitivity.

\noindent{\textbf{{Coding rate. }}} LoRaWAN specifies a coding rate of 4/5 to be used in the US915 band to limit the maximum dwell time in each channel~\cite{LoRaWAN_RP2}. We use a coding rate of 4/5 in both the conventional and offloading-lading mode.

\noindent{\textbf{Spreading factor. }} In the conventional mode, the nodes use spreading factors in the range [9,10] for the US band and [9, 12] for the EU band to enable long-distance communication between nodes and base stations. The ADR algorithm gives a specific spreading factor for each node based on the signal to noise ratio observed at the gateway.  In the offloading mode, depleting nodes use lower spreading factors such as 7 and 8 to communicate with the affluent node. Lower spreading factors typically consume less power and are known to have shorter communication range, and hence, are better suited for packet offloading. Low spreading factors are used only for packet offloading to reduce the number of collisions with other transmissions, thereby conserving energy at the depleting node.  In the lading mode, a node receives packets on low spreading factors but uses high spreading factor to transmit packets to the gateway like conventional mode.

%%% Comment - Potential for increasing the contribution to the paper
%\subsection{Dynamic parameter selection}

\subsection{Energy Overhead during Lading Mode}
\label{sec:lading_energy}
\revise{Typically, the energy consumed by the communication module in an embedded low-power device is higher than that of the computation platform \cite{bandyopadhyay2003energy}. Thus, in this work, we focus on estimating the energy overhead required by the communication module of an affluent node in the lading mode and on ensuring that the overhead does not decrease the lifetime of the affluent node. }  \revise{Note that the energy overhead derived in this section is based on the energy model for communication given in the datasheet of LoRa communication model~\cite{sx1276_datasheet}. In the lading mode, there are three main sources of energy overheads: periodic CAD, packet reception and forwarding, and acknowledgement transmission and reception.}

% and it maintains the accuracy of LoRa energy model presented in the data sheet \cite{sx1276_datasheet}.} 

\noindent\revise{\textbf{Periodic CAD.} Affluent node $v$ in the lading mode performs CAD once every $T_{\text{CAD}} (v)$ seconds, where $T_{\text{CAD}} (v) = T_1 + T_2$ (as shown in Fig. \ref{fig:lading_mode}). Within $T_{\text{CAD}} (v)$ interval, $v$ consumes $P_{rx}$ power for a duration of $\frac{2^{1 + SF_{\text{offload}}}}{BW_{\text{offload}}}$ to probe for correlation. The affluent node consumes $P_\text{RC-Oscillator}$ for a duration of $T_2 - \frac{2^{1 + SF_{\text{offload}}}}{BW_{\text{offload}}}$ to remain in low-power idle mode with only RC-Oscillator on. For the rest $T_1$ time units, the affluent node sleeps and consumes close to $0$ energy. Thus, the total energy consumed during an interval of $T_{\text{CAD}} (v)$ seconds is }
\begin{equation*}
\vspace{-1mm}
%\label{eq:energy_CAD}
%E_{\text{CAD}} (v) =  & T_2 P_{\text{RC-Oscillator}} \\& +  \frac{2^{SF_{\text{offload}}} + 32}{BW_{\text{offload}}} \big( P_{\text{rx}} -  P_{\text{RC-Oscillator}}\big)
E_{\text{CAD}} (v) =   T_2 P_{\text{RC-Oscillator}} +  \frac{2^{( 1 + SF_{\text{offload}})}}{BW_{\text{offload}}} \big( P_{\text{rx}} -  P_{\text{RC-Oscillator}}\big)
%\vspace{-1mm}
\end{equation*}

%\vspace{-3mm}
%\label{eq:energy_CAD}
%\begin{split}
%%E_{\text{CAD}} (v) =  & T_2 P_{\text{RC-Oscillator}} \\& +  \frac{2^{SF_{\text{offload}}} + 32}{BW_{\text{offload}}} \big( P_{\text{rx}} -  P_{\text{RC-Oscillator}}\big)
%E_{\text{CAD}} (v) =  & T_2 P_{\text{RC-Oscillator}} \\& +  \frac{2^{( 1 + SF_{\text{offload}})}}{BW_{\text{offload}}} \big( P_{\text{rx}} -  P_{\text{RC-Oscillator}}\big)
%\end{split}
%\vspace{-2mm}
%\end{equation}

\revise{The actual power consumption values for $P_\text{rx}$ and $P_\text{RC-Oscillator}$ can be obtained from the datasheet of LoRa communication module, and typically power consumed in receive mode is twice that of low-power idle mode \cite{sx1276_datasheet}.}
\revise{Because an affluent node $v$ remains in lading mode for a total duration of $T_{LM}$ time units, the total energy consumed during CAD is given by
$\frac{T_\text{LM} (v)}{T_\text{CAD}} E_\text{CAD} (v)$.} 

%Equation (\ref{eq:energy_CAD}) gives the energy overhead of affluent node $v$ during CAD operation for a lading period of $T_{\text{CAD}} (v)$ seconds. In the interval of $T_\text{CAD}$, an affluent node is active for $T_2$ seconds and sleeps for $T_1$ seconds. Of the $T_2$ seconds, an affluent node listens for packet headers for two symbols, i.e., the ratio of ${2^{( 1 + SF_{\text{offload}})}}$ and ${BW_{\text{Offload}}}$ and is idle with RC oscillator for the rest. Here, $SF_{\text{offload}}$ and ${BW_{\text{Offload}}}$ represent the transmission parameters (spreading factor and bandwidth, respectively) used to offload a packet from $u$ to $v$. When a node is idle with RC oscillator ON, it consumes $P_{\text{RC-Oscillator}}$ power.  For listening, a node consumes $P_{\text{rx}}$ power. On average, the power consumption for receiving a packet is double the power consumption in idle mode with RC oscillator ON.

\noindent \revise{\textbf{Packet reception and forwarding.} The total energy overhead to receive a packet in the lading mode is a product of the power drawn by the chip and time on air of the packet, which is given by the product of number of symbols in the packets and time to transmit one symbol. In the lading mode, a node receives packets transmitted to it using the offloading-lading transmission parameters. The number of symbols in $u$'s packet when transmitted by $u$ to $v$ using the offloading-lading transmission parameters, is given as follows:}
\vspace{-2mm}
\begin{equation}
\label{eq:number_symbols_u_v}
\begin{split}
 & N_{\text{symbols}} ( u \xrightarrow{u} v) = \text{preamble}_u + 4.25 + 8 \\& + \max \bigg( \bigg\lceil  \frac{8 \, \text{payload}_u - 4\,SF_{\text{offload}} + 24}{SF_{\text{offload}} - 2\, \text{DE}_{\text{offload}}}  \bigg\rceil \frac{1}{CR_{\text{offload}}}, 0 \bigg)
\end{split}
\end{equation}
\revise{Similarly, the time to transmit one symbol of node $u$'s packet from $u$ with offloading-lading transmission parameters is:}
\vspace{-2mm}
\begin{equation}
\label{eq:length_symbols_offloading}
T_{\text{symbol}} (u \xrightarrow{u} v) = \frac{2^{SF_{\text{offload}}}}{BW_{\text{offload}}}
\vspace{-1mm}
\end{equation}
\revise{To receive a packet from $u$, the affluent node $v$ consumes $P_{\text{rx}}$ power for a duration of $N_{\text{symbols}} (u \xrightarrow{u} v)  T_{\text{symbol}} (u \xrightarrow{u} v)$. }

\revise{In the lading mode, the affluent node forwards offloaded packet using its conventional mode parameters. Thus, the number of symbols in $u$'s packet when transmitted by $v$ using $v$'s conventional mode transmission parameters is given by}
\vspace{-2mm}
\begin{equation}
\label{eq:number_symbols_v_gateway}
\begin{split}
& N_{\text{symbols}} (v \xrightarrow{u} \text{Gateway}) = \text{preamble}_u + 4.25 + 8 \\& + \max \bigg( \bigg\lceil  \frac{8 \, \text{payload}_u - 4\,SF_{\text{v}} + 24}{SF_{\text{v}} - 2\, \text{DE}_{\text{v}}}  \bigg\rceil \frac{1}{CR_{\text{v}}}, 0 \bigg)
\end{split}
%\vspace{-0.5mm}
\end{equation}
\revise{Similarly, the time to transmit on symbol of node $u$'s packet from node $v$ to Gateway with $v$'s conventional mode transmission parameters is given by Equation (\ref{eq:symbol_duration}). To forward an offloaded packet, an affluent node $v$ consumes $P_{\text{tx}} (PL_v)$ power for $N_{\text{symbols}} (v \xrightarrow{u} \text{Gateway}) T_{\text{symbol}} (v)$ duration. Thus, the total energy consumed by an affluent node $v$ for receiving and forwarding one packet from $u$ is given as follows:}
\vspace{-2mm}
\begin{equation}
\vspace{-2mm}
\label{eq:energy_forwarding}
\begin{split}
& E_{\text{FW}}(u, v) =  P_{\text{rx}} \, N_{\text{symbols}} (u \xrightarrow{u} v) \, T_{\text{symbol}} (u \xrightarrow{u} v)  \\& + P_{\text{tx}} (PL_v) \, N_{\text{symbols}} (v \xrightarrow{u} \text{Gateway}) \, T_{\text{symbol}} (v) 
\end{split}
%\vspace{-0.5mm}
\end{equation}

\revise{Note that the number of symbols and time to transmit a packet using the offload parameters, as given by Equation (\ref{eq:number_symbols_u_v}) and (\ref{eq:length_symbols_offloading}) are significantly lower than that using the conventional parameters, as given by Equation (\ref{eq:number_symbols_v_gateway}) and (\ref{eq:symbol_duration}). Furthermore, the power level used for transmission in offloading-lading mode is lower than that of conventional mode. Thus the total energy consumed during a packet's offloading is significantly lower than that needed for its transmission to the gateway.}

\noindent \revise{\textbf{Acknowledgment transmission and reception.} Similar to packet transmission, energy overhead during transmitting and receiving acknowledgments is a product of power drawn, number of symbols, and time to transmit one symbol. In the lading mode, a node transmits an ACK to a depleting node using the offloading-lading transmission parameters. Thus, the number of symbols in node $v$'s ACK to node $u$ is given by Equation (\ref{eq:number_symbols_v_u}), assuming that the packet transmitted contains only the preamble and header (i.e., no payload and CRC).}
\vspace{-1mm}
\begin{equation}
\label{eq:number_symbols_v_u}
\begin{split}
&N_{\text{symbols}} (v \xrightarrow{\text{ack}} u) = \text{preamble}_u + 4.25 + 8 + \\& \max \bigg( \bigg\lceil  \frac{8 - 4\,SF_{\text{offload}}}{SF_{\text{offload}} - 2\, \text{DE}_{\text{offload}}}  \bigg\rceil \frac{1}{CR_{\text{offload}}}, 0 \bigg)
\end{split}
\end{equation}
\vspace{-1mm}
\revise{Here, $v \xrightarrow{\text{ack}} u$ denotes the transmission of ACK from node $v$ to node $u$ after a successful packet reception.
The duration of each symbol in node $v$'s ACK to node $u$ ($T_{\text{symbol}} (v \xrightarrow{\text{ack}} u)$) is the same as $T_{\text{symbol}} (u \xrightarrow{u} v)$ since the transmission parameters are the same. Equation (\ref{eq:length_symbols_offloading}) gives the expression for $T_{\text{symbol}} (u \xrightarrow{u} v)$. 
The affluent node transmits an acknowledgment twice and each acknowledgment takes $N_{\text{symbols}} (v \xrightarrow{ack} u) T_{\text{symbol}} (u \xrightarrow{u} v)$  time to transmit and consumes $P_{\text{tx}} (PL_{\text{offload}})$ power. In the lading node, node $v$ receives an ACK for node $u$'s packet from the gateway using its transmission parameters. To receive an ACK, node $v$ opens two receive window slots with a total length of $T_\text{rx} (v)$. For the  duration of $T_\text{rx}$, it consumes $P_\text{rx}$ power. Thus, the total energy consumed to transmit and receive an ACK is:}
\vspace{-1mm}
\begin{equation}
\vspace{-1mm}
\label{eq:energy_ack}
\begin{split}
& E_{\text{ack}}(u, v) = 2P_{\text{rx}} \,  T_\text{rx} (v) \\& +  2P_{\text{tx}} (PL_{\text{offload}}) \, N_{\text{symbols}} (v \xrightarrow{\text{ack}} u) \,T_{\text{symbol}} (u \xrightarrow{u} v)
\end{split}
\end{equation}
%\vspace{-2mm}

\revise{A depleting node $u$ generates $\frac{T_{\text{LM}} (v)}{\tau_u}$ packets in the lading mode of $v$, where $\tau_u$ is its minimum inter-packet generation time. If each packet of node $u$ needs an average of $\gamma$ retransmissions to successfully reach $v$, the total energy consumed to offload $\frac{T_{\text{LM}} (v)}{\tau_u}$ packets including ACKs is given by}
\vspace{-1mm}
$$ \vspace{-1mm}\frac{\gamma T_{\text{LM}} (v)}{\tau_u} \big(  E_{\text{FW}} (u,v) + E_{\text{ack}} (u,v) \big)$$

A node does not perform CAD during the reception and transmission of a packet or an ACK. Thus, the energy correction is the product of $E_{\text{CAD}} (v)$ and the number of occurrences of CAD during the reception and transmission of both the offloaded packet and its ACK. The energy correction at node $v$ per packet offloaded by $u$ is given by Equation (\ref{eq:energy_correction}).
\vspace{-1mm}
\begin{equation}
\vspace{-1mm}
\label{eq:energy_correction}
\begin{split}
& E_{\text{CR}}(u,v) =  \frac{E_{\text{CAD}}(v)}{T_{\text{symbol}}}  \bigg( 2T_{rx} \\& + 2N_{\text{symbols}} (v \xrightarrow{\text{ack}} u) \,T_{\text{symbol}} (u \xrightarrow{u} v) \\& +  N_{\text{symbols}} (u \xrightarrow{u} v) \, T_{\text{symbol}} (u \xrightarrow{u} v) \\& + N_{\text{symbols}} (v \xrightarrow{u} \text{Gateway}) \, T_{\text{symbol}} (v) \bigg)
\end{split}
\end{equation}
\vspace{-1mm}

\noindent \revise{\textbf{Overall overhead. } Considering packet offloading from multiple depleting nodes and three sources of energy overhead, the total energy overhead in the lading mode of a node  $v$ is given by Equation (\ref{eq:energy_lading}), where $\text{DN}$ represents the set of depleting nodes, and $\text{AF\_DN}$ represents the set of affluent and depleting node pairs that communicate with each other.}
\vspace{-1mm}
\begin{equation}
\vspace{-1mm}
\label{eq:energy_lading}
\begin{split}
&E_\text{LM} (v) = \frac{T_\text{LM} (v)}{T_\text{CAD}} E_\text{CAD} (v) + \\&\sum_{\substack{ \{u, v\} \in \text{AF\_DN} \\ u \in \text{DN}}} \frac{\gamma \, T_\text{LM} (v)}{\tau_u}\big( E_{\text{FW}}(u,v) + E_{\text{ack}} (u,v) - E_{\text{CR}}(u,v) \big)
\end{split}
 %\vspace{-0.1mm}
\end{equation}
 Note that Equation (\ref{eq:energy_lading}) gives an upper bound on the energy overhead on the lading mode, assuming that nodes are always performing CAD during the lading mode. In practice, the nodes can be sleeping for short periods, i.e., between inter-arrival times of packets. To simplify the computation of affluent and depleting node pairs that communicate with each other, we use this upper bound on the lading mode's energy consumption. 
 \edit{As we explain in the following section, this energy overhead for an affluent node is used to find the affluent and depleting node pairs that communicate with each other.}
%\revise{This energy overhead for an affluent node is used to find the affluent and depleting node pairs that communicate with each other. Selection of affluent-depleting nodes pairs and the duration of operation is discussed in the following section.}

%\input{energyModel}
%!TEX root = main.tex

\subsection{Selecting Different Modes of Operation}
\label{sec:selectingModes}

%The base station predicts the traffic pattern through each node with the help of ML techniques. Various machine learning techniques have been proposed to accurately predict the wireless network traffic in literature. Examples include the Autoregressive moving integrated average technique (ARIMA), Seasonal ARIMA~\cite{SARIMA}, Support Vector Machines, Multi-layer perception,~\cite{MLP} Gaussian Process and Long Short Term Memory (LSTM) architecture~\cite{LSTM}. We will use the LSTM based method as it has shown to be effective in predicting time series data such as network traffic. We will discuss the LSTM algorithm in detail in subsection~\ref{sec:LSTM}. 

\subsubsection{Workload Prediction}

%%why we need prediction
The network server distinguishes affluent nodes from depleting nodes based on the total energy consumption of the node in the current recharge cycle. Unfortunately, the unreliability in wireless networks makes it difficult to predict the number of transmission attempts each node will make before the packet has been successfully received at the gateway. In case of packet reception failure at the gateway, a node can retransmit it up to $8$ times. However, the actual number of retransmissions can be less than that. 
%%how do we predict in our approach 

To estimate the number of retransmissions from the nodes, we use a sliding window approach. Each node embeds the retransmission attempt information in the packet it transmits. The network server keeps the retransmission count from the last $w$ packets in a sliding window. The size of the sliding window can be decided by the network manager. The average retransmission count of the window is used as the retransmission overhead, $\gamma$, for the network. There are existing works that use various machine learning techniques such as Auto-Regressive Moving Integrated Average Technique (ARIMA), Seasonal ARIMA~\cite{SARIMA}, Support Vector Machines, Multi-Layer Perception,~\cite{MLP} Gaussian Process and Long Short Term Memory (LSTM) architecture~\cite{LSTM} to predict the wireless network traffic pattern. However, such approaches increase the complexity of the system. Thus, we leave the integration of such prediction algorithms in our approach as a future work.

%%question: Do we use gateway of network server in the paper

\subsubsection{Selection of Affluent and Depleting Nodes}
%%First algorithm to decide two sets affluent and depleting
The network server selects the affluent and depleting nodes based on their expected energy consumption in the conventional mode. For each node in $v$ in the network, it estimates the  total number of transmission attempts $N_\text{tx}(v)$ in the current recharge cycle as  $ N_\text{tx} (v) = \gamma \times \frac{T_\zeta}{\tau_v}$, 
%\vspace{-2mm}
%\begin{equation}
%    N_\text{tx} (v) = \gamma \times \frac{T_\zeta}{\tau_v}
%\end{equation}
%\vspace{-.5mm}
where $T_\zeta$ denotes the time interval from current time $t$ to the start of the next recharge cycle. Using this information, the network server estimates the energy consumption in conventional mode for each node $v$ as  $E_\text{CM}(v) = N_\text{tx} \times E_\text{tx} (v)$, 
%\vspace{-1.5mm}
%\begin{equation} \label{eq:energy_conv}
%   E_\text{CM}(v) = N_\text{tx} \times E_\text{tx} (v) 
%\end{equation}
%\vspace{-0.5mm}
where,  $E_\text{CM}(v)$ is the energy consumption in conventional mode and $E_\text{tx} (v) $ is the energy consumption for packet transmission given by Equation (\ref{eq:energy_transmission}). The network server can estimate the battery budget $B_v^*$ for each node $v$ in the network. Thus, if $B_v^* -\beta_v > E_\text{CM} (v)$, it assigns node $v$ to the set of affluent nodes $\text{AN}$. Otherwise, we assign node $v$ to the set of depleting nodes $\text{DN}$. %Here, $\beta_v$ is the amount of budget that is saved for the next interval. 
In this way, the network server divides all nodes into two disjoint sets $\text{AN}$ and $\text{DN}$ as presented in Algorithm~\ref{alg:select_lading_offloading_1}. \revise{The server runs this Algorithm after every packet reception.}
%as soon as it detects a change in $E_\text{CM}(v)$ for any node $v$ in the network. 
%%question how often it is executed, after every heartbeat message/ after the retransmission overhead is changed

%\vspace{2mm}
\begin{algorithm}[!t]
\scriptsize
\caption{Selecting Affluent and Depleting Nodes}\label{alg:select_lading_offloading_1}

\begin{algorithmic}
\STATE \textbf{Input:} Set of all nodes $V$  
\STATE \textbf{Output:} Disjoint sets $\text{AN}, \text{DN}$ 
\STATE $AN \gets \emptyset$
\STATE $DN \gets \emptyset$
\FOR {$v \in V$} 
        \STATE $E_\text{CM}(v)  \gets N_\text{tx} \times E_\text{tx} (v)$
        \IF {$E^\text{CM} (v)  < B_v^* -\beta_v $}
                \STATE $AN \gets AN \cup \{v\}$
        \ELSE
                \STATE $\text{DN} \gets \text{DN} \cup \{v\}$
        \ENDIF
\ENDFOR
\end{algorithmic}
\end{algorithm}
%\vspace{-2mm}

\subsection{Selecting Lading and Offloading Mode}

After executing Algorithm~\ref{alg:select_lading_offloading_1}, if the network server identifies any depleting node $u \in \text{DN}$, it immediately starts to look for a suitable affluent node $v \in \text{AN}$ in the same cell which can support the lading mode of operation. To this end, the network server creates the set $\text{AF\_DN}$ of affluent and depleting nodes that communicate with each other. Each member in the set $\text{AF\_DN}$ is a pair $(u,v)$ where $u$ is the depleting node and $v$ is the affluent node.
%After such a pair of affluent and depleting nodes $(u,v)$ is selected,  
For every pair $(u,v) \in \text{AF\_DN}$, the lading time $T_\text{LM} (v) $ for affluent node $v$ should be at least as large as $\tau_{u}$, one sampling period of $u$, to successfully enable offloading of at least one packet from $u$ to $v$. Thus, every pair $(u,v) \in \text{AF\_DN}$  should satisfy the constraint: 
\vspace{-2mm}
\begin{equation}
\vspace{-2mm}
 \label{eq:time constraint}
 T_\text{LM} (v) \geq \tau_{u} \hspace{5pt} \forall (u,v) \in \text{AF\_DN}.  
\end{equation}
Furthermore, in order to prevent depletion of budget at the affluent node, the time for lading mode should be carefully selected so that energy consumption in lading mode does not exceed the \redundantResidualEnergy $E_r$ for $v$, calculated as the difference between node $v$'s budget and $E_\text{CM} (v) $. Thus, for an affluent node $v$, the energy consumption in lading mode $E_\text{LM}$ is bounded as 
\vspace{-4.5mm}
\begin{equation} \label{eq:energy_constraint}
E_\text{LM} (v) \leq E_{r} (v).
\end{equation}
%\vspace{-0.5mm}

Since optimal pairs of affluent and depleting nodes are unknown, we rely on a greedy heuristic which enables an affluent node to assist as many depleting nodes as possible without hampering its own lifetime. To this end, the network server constructs the set $\text{AF\_DN}$ by pairing an affluent node $v$ with every depleting node $u$ in the same cell. Then, the network server calculates the time in lading mode $T_\text{LM} (v)$ such that Equation (\ref{eq:energy_constraint}) is satisfied. \revise{Such an approach prevents an affluent node from becoming a depleting node.} Note that, if node $v$ is paired with a large number of depleting nodes, $T_\text{LM} (v) $ is shortened to keep $E_\text{LM} (v) $ bounded. Thus, the estimated value of  $T_\text{LM}$ might not satisfy Equation (\ref{eq:time constraint}). In other words, the affluent node may not be able to support packet \edit{offloading from these many depleting nodes $u \in \text{AF\_DN}$ with its superfluous residual energy.}  Thus, the network server needs to remove some pairs from $\text{AF\_DN}$ to increase $T_\text{LM} (v) $ and satisfy Equation (\ref{eq:time constraint}). Since our goal is to prolong the lifetime of depleting nodes, we select to remove the pairs with the highest \textit{offloading overhead} $E_\text{OM}(u,v)$ given by the following equation which represents the energy consumption for node $u$ to offload one packet to node $v$: 

\vspace{-4mm}
{\small
\begin{equation*} %\label{eq:energy_offload}
    E_\text{OM}(u,v) = P_{\text{tx}} (PL_\text{OFFLOAD}) \times  N_{\text{symbols}} ( u \xrightarrow{u} v) \; T_{\text{symbol}} (u \xrightarrow{u} v)
    \vspace{-5mm}
\end{equation*}
}

%\vspace{-1mm}
In the above equation, $PL_\text{OFFLOAD}$ is the transmission power for node $u$ in offloading mode. Thus, the network server greedily removes the pair $(u,v)$ with the highest offloading overhead from the set $\text{AF\_DN}$ until node $v$ is able to support packet offloading for every pair $(u,v)$ in the set as described in Algorithm~\ref{alg:select_lading_offloading_2}. After the set $\text{AF\_DN}$ is constructed, for each pair $(u,v) \in \text{AF\_DN}$, the network server enables offloading and lading mode in $u$ and $v$, respectively, for the duration of $T_\text{LM} (v)$. Note that, the network server only provides an upper bound of time in lading and offloading mode.
%$T_\text{LM} (v)$ is large enough to support packet offloading from all pairs. 
%The process is described in Algorithm~\ref{alg:select_lading_offloading_2}. 
%algorithm to decide offloading nodes to lading nodes here

\begin{algorithm}[!t]
\scriptsize
\caption{Selecting Lading and Offloading Mode}\label{alg:select_lading_offloading_2}
\begin{algorithmic}
\STATE \textbf{Input:} Disjoint sets $\text{AF}, \text{DN}$ 
\STATE \textbf{Output:} Set $\text{AF\_DN}$  

\FOR {$v \in \text{AN}$} 
        \STATE $\text{AF\_DN} \gets \{ v \} \times \text{DN}$
        \FOR {$(u,v) \in \text{AF\_DN}$}
      %  \IF {$C_{i}\neq C_{j}$}
      %          \STATE $\text{OM}(v_i) \gets \text{OM}(v_i) \setminus \{v_j\}$
      %  \ENDIF
      %  \IF{$E^\text{OM}_{i,j} \ge E^\text{tx}_{j}$}
      %          \STATE $\text{OM}(v_i)\gets \text{OM}(v_i)\setminus \{v_j\}$
      %  \ENDIF
       %
       \STATE sort $\text{AF\_DN}$ in ascending order of $E_\text{OM} (u,v)$
       \STATE \text{calculate} $T_{LM} (v) $ \text{for} $\text{AF\_DN}$ \text{from Equation (\ref{eq:energy_lading}})
      % \WHILE{$T_{i}^{LM} \leq T_{j} \hspace{5pt} \forall v_j \in \text{OM}(v_i)$}
       \WHILE{Equation~\ref{eq:time constraint} is not satisfied}
       \STATE  \text{Remove last pair} $(u,v) $ \text{from} $\text{AF\_DN}$
      % \STATE $\text{OM}(v_i)\gets \text{OM}(v_i) \setminus v_k$
       \STATE \text{calculate} $T_\text{LM} (v) $ \text{for} $\text{AF\_DN}$ \text{from Equation (\ref{eq:energy_lading}})
       \ENDWHILE 
       \ENDFOR
\ENDFOR
\end{algorithmic}
\end{algorithm}

%\edit{If an affluent node $v$ consumes more energy in offloading packets than estimated by the network server, it can stop lading mode and start operating in the conventional mode. }

\subsubsection{Transitioning between Conventional and Lading/Offloading Modes}
\label{sec:mac_transition}
Upon deciding the set $\text{AF\_DN}$, the network server broadcasts the information via acknowledgments. One challenge in transitioning from conventional mode to lading/offloading mode is estimating the start of the lading/offloading mode. Since nodes transmit a heartbeat message periodically, the network server can predict the arrival of the next heartbeat message from each node that is transitioning from conventional mode to lading/offloading mode. This information can be used to estimate the next time instant before which all nodes will have a minimum of one communication (either heartbeat or data) with the gateway. The network server chooses this instant as the start of the lading and offloading mode of operation. This ensures that all nodes that are transitioning to offloading/lading modes receive information about the transition.

The network server chooses the affluent node to start its lading mode a few seconds before the offloading mode starts. Such a proactive approach circumvents any time synchronization errors between the devices. When its reserve energy is exhausted, an affluent node  can use piggy-backed acknowledgments to inform the depleting nodes to transition from offloading mode to conventional mode. Once all depleting nodes are notified to switch to the conventional mode, an affluent node can transition to conventional mode.

%!TEX root = main.tex

%\vspace{-2mm}
\section{Evaluation}\label{sec:evaulation}
%\vspace{-2mm}
We have evaluated our results in simulation and experiment.
%We have evaluated the proposed protocol for long-lived LoRa through simulations and proof-of-concept experiments. We first present our large-scale evaluation in simulation. Then we present our experiments. 

\subsection{Simulation}
\subsubsection{Simulation Setup} \label{sec:sim_setup}
We evaluate the efficiency of our approach by conducting simulations using up to 1200 nodes and a single gateway placed in a disc of radius \revise{3500m}. Our implementation is based on the LoRaWAN module for NS-3 \cite{magrin2019thorough}. The nodes and the gateway use 8 channels between 902.3 and 903.7 MHz with a channel bandwidth of 125 kHz. To conform with the US regulations, we do not use SF11 and SF12. Unless stated otherwise, the parameters were set using our transmission parameter selection approach. 

\revise{We configure a  percentage of nodes to transmit at a higher rate of 20 to 30 packets per hour. All other nodes transmit at a rate of 2 to 4 packets per hour. The packet interarrival rate remained unchanged for the duration of the simulation. For each node, a rechargeable battery was simulated by assigning battery budgets randomly in the range [6,25] joules for the current recharge cycle.
%Note that, battery budget is defined as an estimation of the energy generated during one recharge cycle. 
The heterogeneity of battery budgets and traffic patterns of the nodes introduce depleting nodes in the network. 
%Note that, packet offloading is only used if any depleting node exist in the network.
Affluent and depleting nodes, if any, were selected based on Algorithm~\ref{alg:select_lading_offloading_1} after every packet reception at the gateway. For simplicity, we use a fixed retransmission overhead, $\gamma=2$. Energy consumption for packet transmission (including retransmissions), packet reception and CAD are subtracted from this budget. Packet transmission from a node was stopped upon depletion of its budget. 
For each setup, we reported the network lifetime and throughput for a single run. Network throughput was measured by the number of bytes received per unit of time at the gateway.}

\noindent \revise {\textbf{Baseline.} To our knowledge, no existing work studied lifetime maximization of energy-harvesting LoRa networks. Thus we compare the performance of our approach (shown as `LLL'  i.e., Long-Lived LoRa in figures) with standard LoRaWAN.}

\subsubsection{Results under Varying Number of Nodes}
\begin{figure}[ht]
	\centering
	\vspace{-0.2in}
	\subfigure[Network lifetime]{
	\includegraphics[width=0.23\textwidth]{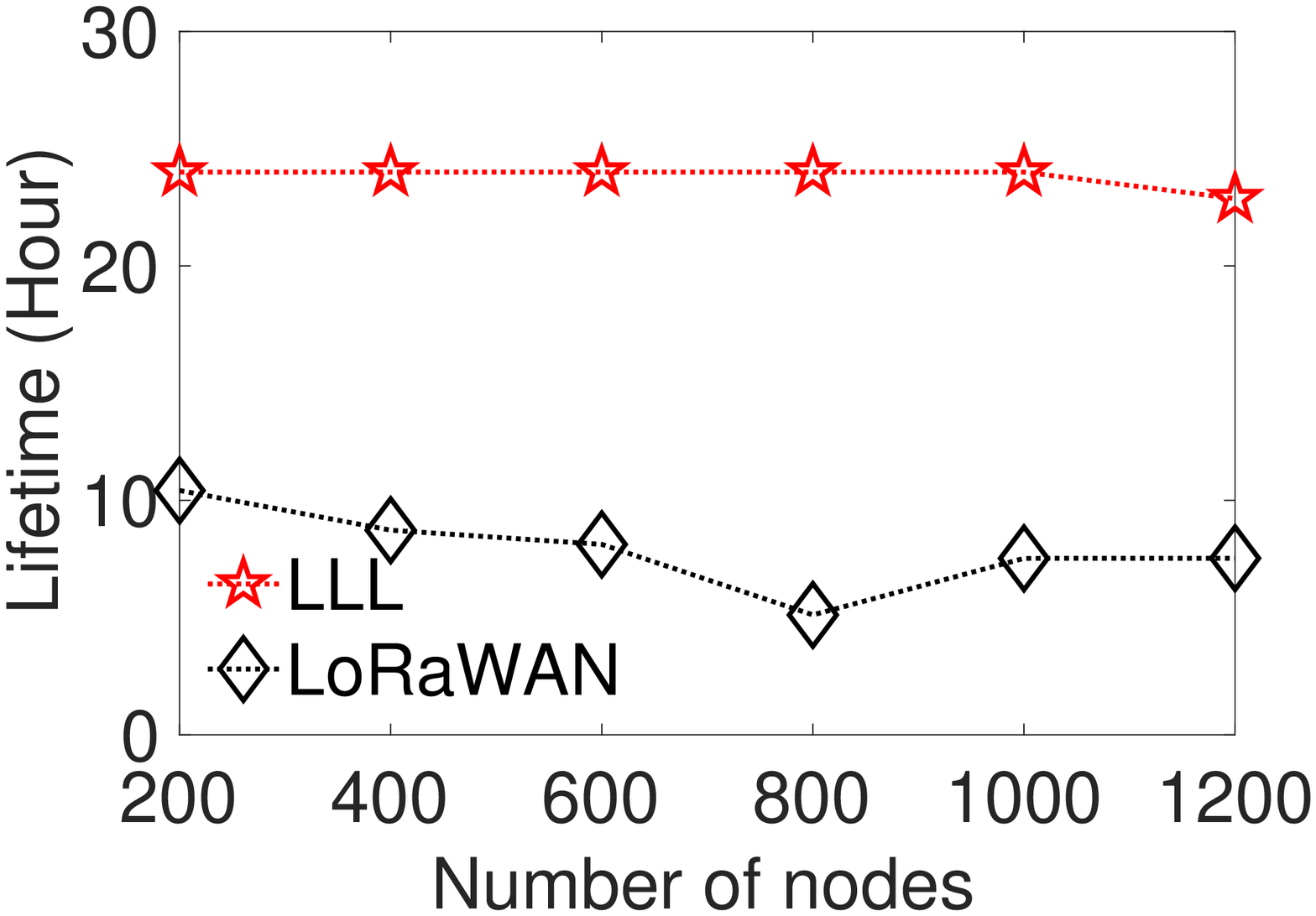}
    \label{fig:vary_nodes_lifetime}
    }%\vfill \vspace{-0.1in}
    \subfigure[Throughput]{
		\includegraphics[width=0.23\textwidth]{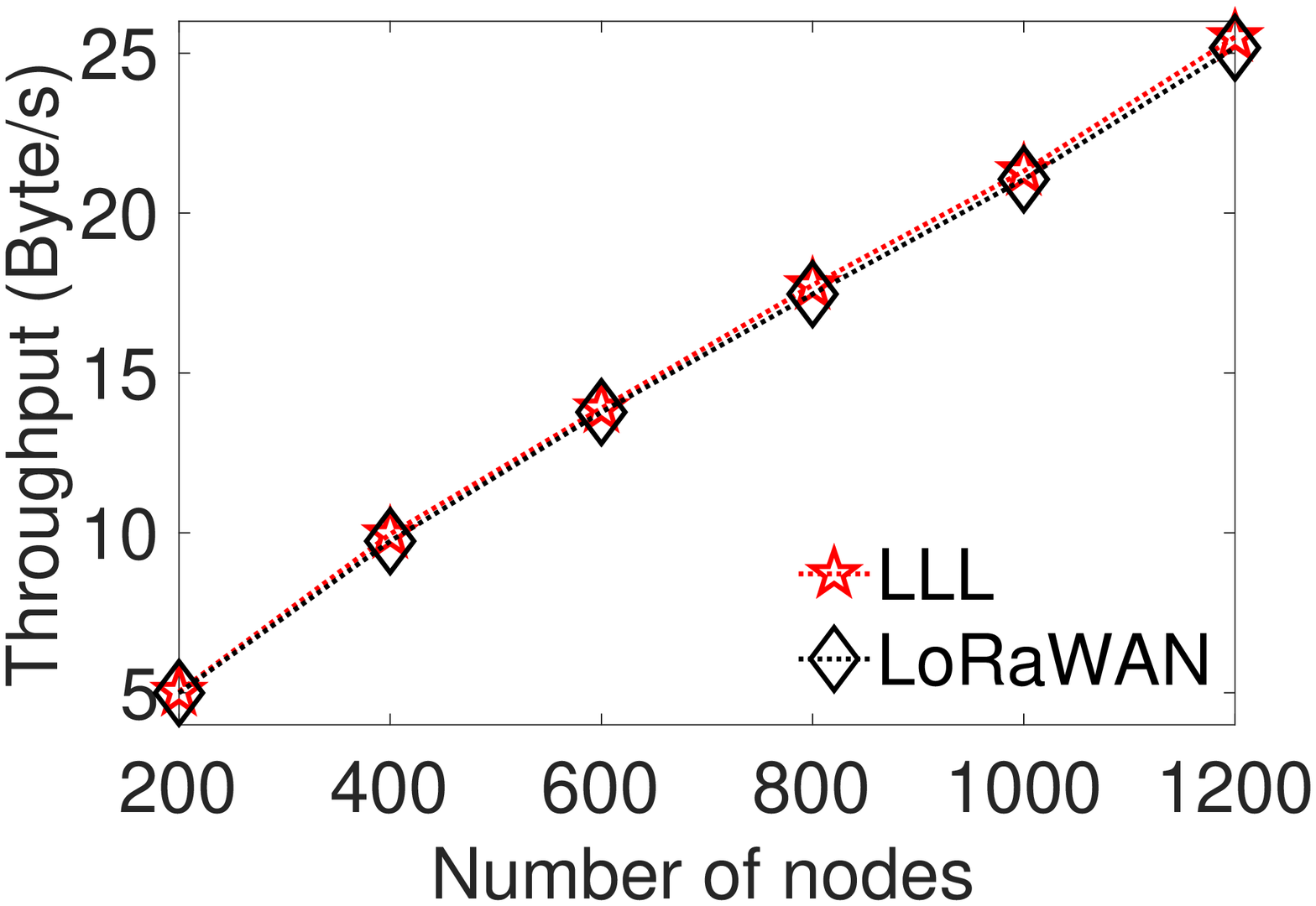}
         \label{fig:vary_node_throughput}
	}%\quad
	\vspace{-0.1in}
	 \caption{Network lifetime and throughput under varying number of nodes}
	\label{fig:vary_nodes}
	\vspace{-0.2in}
\end{figure}

We first evaluate our approach under varying number of nodes, where approximately 3\% of the nodes are depleting. Figure~\ref{fig:vary_nodes} demonstrates the performance in terms of lifetime and throughput. \revise{In Figure~\ref{fig:vary_nodes_lifetime}, the network lifetime remains fairly constant for our approach despite the increase in number of nodes whereas for LoRaWAN the lifetime decreases as the number of nodes increase (at 200 nodes the lifetime is 10.42 Hours for LoRaWAN which decreases to 7.53 Hours at 1200 nodes).
%The behavior is expected because network lifetime is measured according to the first node that fails, which has little dependence on the total number of nodes.  
The small variations in network lifetime across experiments in Figure~\ref{fig:vary_nodes_lifetime} are mainly due to the randomly assigned battery budgets. As the figure shows, our approach (LLL) is consistently better than the legacy LoRa (LoRaWAN). Our approach has a lifetime of 23.8 Hours on average compared to 7.53 Hours in LoRa. When the number of nodes is 800, we achieve more than 4 times better performance in terms of lifetime. As shown in Figure \ref{fig:vary_node_throughput},  LLL maintains throughput no less than the standard LoRaWAN under the same set up.} 

%
%As the number of nodes increases, the network throughput of both approaches increases (Figure \ref{fig:vary_node_throughput}). On average the network throughput of our approach is 15.5 Byte/s while for LoRa it is 15.36 Byte/s.  Overall, Figure~\ref{fig:vary_nodes} shows that our approach increases the lifetime significantly while maintaining throughput no less than LoRaWAN. 

\revise{It is worth noting that, in our approach, offloading may increase the latency of some (offloaded) packets. However, the applications deployed in a LoRa network such as data collection in smart agriculture~\cite{agricultureIoT}, smart metering~\cite{smart_meter}, smart cities~\cite{airIoT} and wild-life monitoring~\cite{wildlife} are usually not time-sensitive. Rather, lifetime maximization is of greater importance as it ensures seamless data collection from all sensors. Thus, by increasing the latency of some packets, we are significantly prolonging the lifetime of the network while keeping the throughput no less than the standard LoRaWAN.} 
%As a result, the behavior of the applications analyzing those collected data at the gateway remains mostly unaffected because the time to collect a fixed number of packets from the nodes remains almost unchanged.
%For a network with 400 nodes using 8 channels and single gateway, we see 3 times improvement in the network lifetime and 21\% increase in network throughput compared to traditional LoRaWAN.

%\subsubsection{Varying number of nodes}
%\begin{figure}
%    \centering
%    \includegraphics[width=0.45\textwidth]{Fig/varyNodes_lifetime.eps}
%%    \caption{Network lifetime under varying number of nodes}
%    \label{fig:vary_nodes_lifetime}
%\end{figure}

%\begin{figure}
 %   \centering
%    \includegraphics[width=0.45\textwidth]{Fig/varyNodes_throughput.eps}
%    \caption{Throughput under varying number of nodes}
%    \label{fig:vary_node_throughput}
%\end{figure}

\subsubsection{Results under Varying Number of Depleting Nodes}
\begin{figure}[ht]
	\centering
	\vspace{-0.2in}
	\subfigure[Network lifetime]{
	\includegraphics[width=0.23\textwidth]{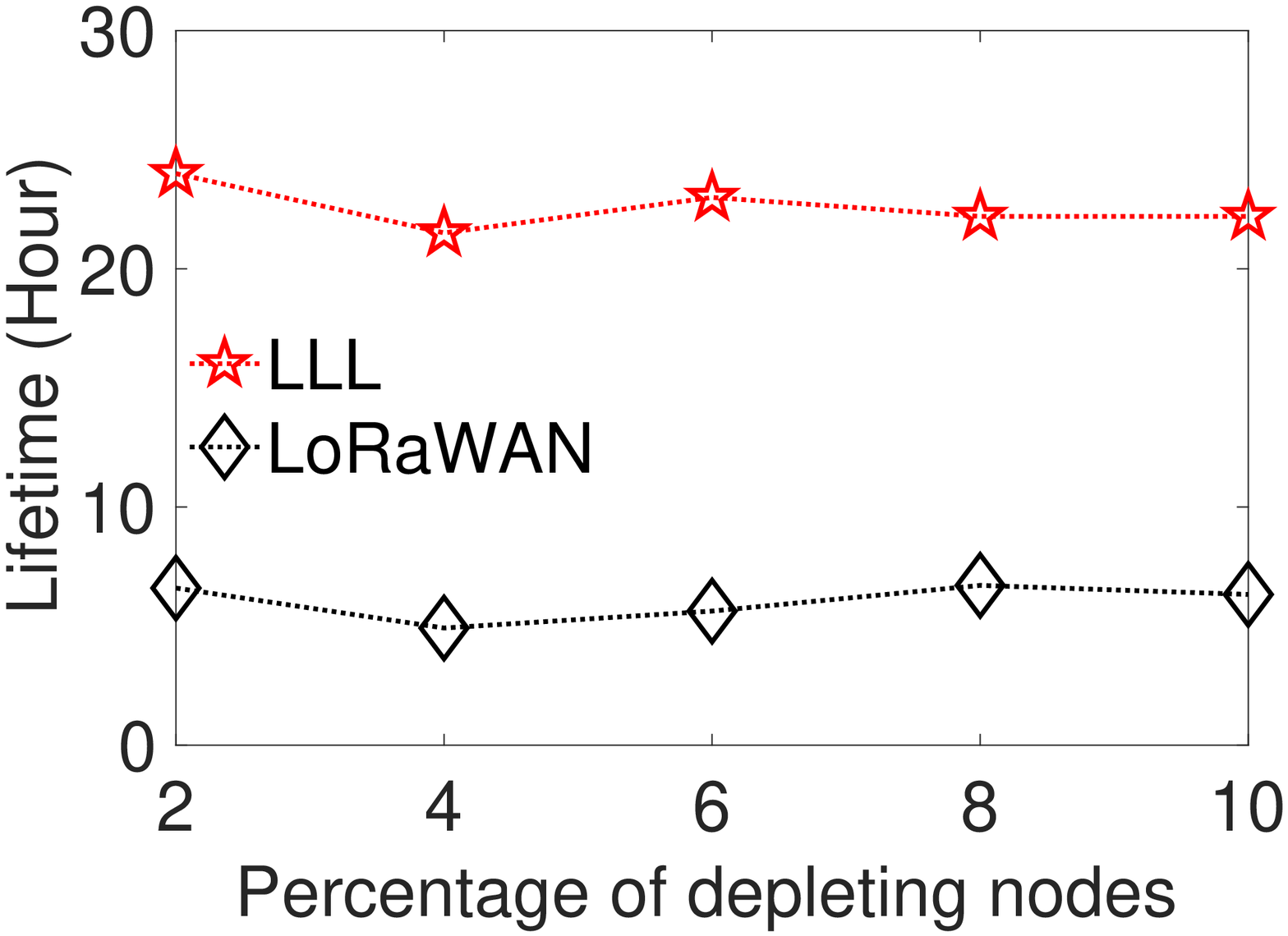}
    \label{fig:vary_off_lifetime}
    }%\vfill \vspace{-0.1in}
    \subfigure[Throughput]{
		\includegraphics[width=0.23\textwidth]{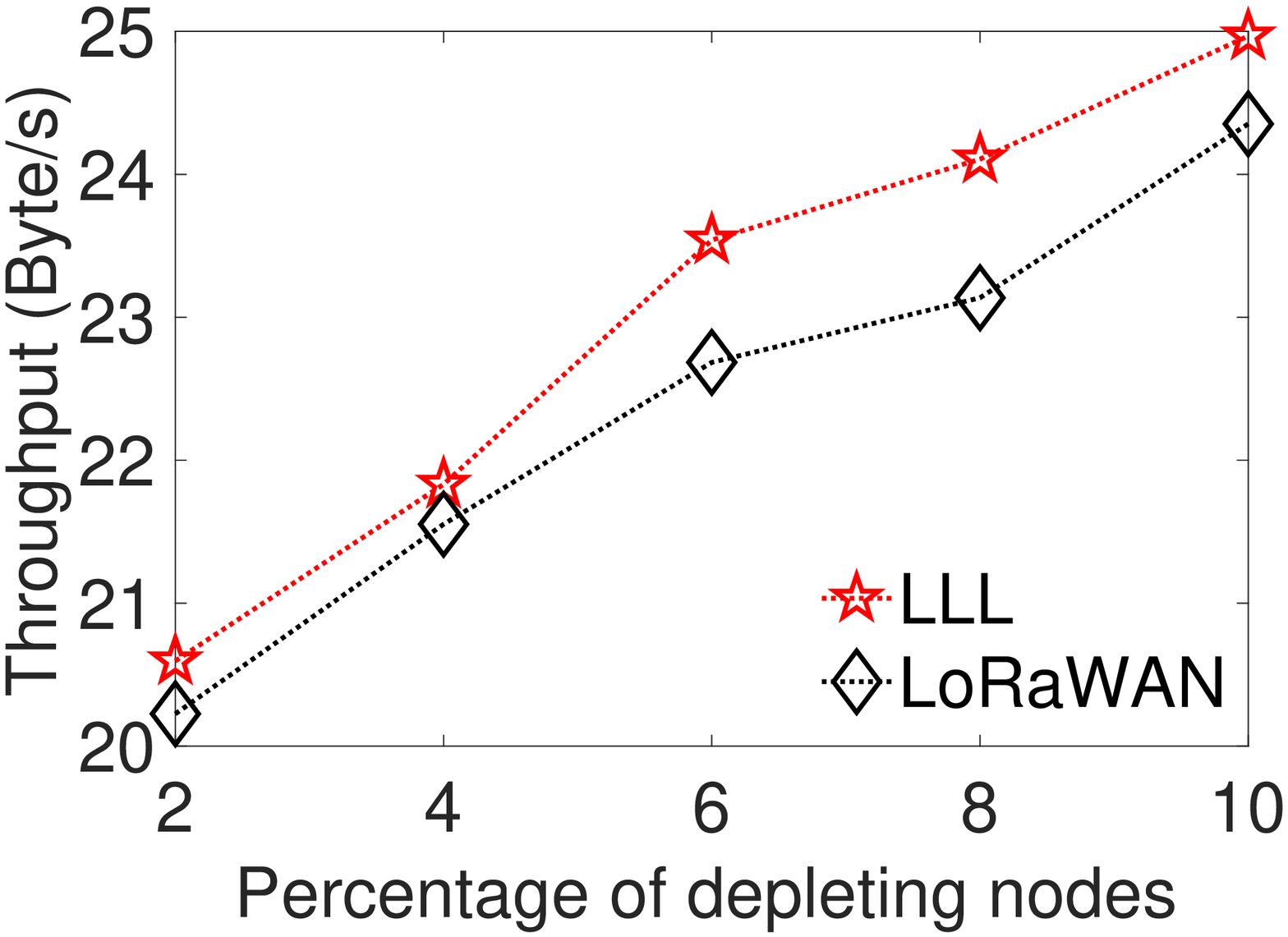}
         \label{fig:vary_off_throughput}
	}%\quad
	\vspace{-0.1in}
	 \caption{Lifetime and throughput under varying \% of offloading nodes}
	\label{fig:vary_off}
	\vspace{-0.1in}
\end{figure}

We now evaluate our approach under varying percentage of depleting nodes. We fix the number of nodes at 1000 and vary the percentage of depleting nodes in the network from 2 to 10. As shown in Figure~\ref{fig:vary_off_lifetime}, the lifetime of our approach is significantly better than the traditional LoRa, as the percentage of the depleting nodes increases. \revise{On average, the network lifetime for our approach is 22.58 Hours, whereas for LoRaWAN, the lifetime is 6.03 Hours. We observe the lifetime for our approach decreases as the percentage of depleting node increase as the set of affluent nodes are not able to support a large number of depleting nodes (at 2\% depleting nodes, the lifetime is 24 Hours which decreases to 22 Hours at 10\% depleting nodes). Figure~\ref{fig:vary_node_throughput} shows that our approach has network throughput of 23 Byte/s where LoRaWAN has 22.39 Byte/s. As we introduce more depleting nodes in the network by increasing the packet rate at some nodes, the throughput of both approaches increase. However, our approach prevents early depletion of batteries and thus achieves higher throughput than  LoRaWAN under many depleting nodes.} 
%Overall, our simulation results show that our approach significantly outperforms LoRa in terms of lifetime in large-scale deployment.
\subsubsection{Results under Duty Cycle Enabled EU Band}
\begin{figure}[ht]
	\centering
	\vspace{-0.25in}
	\subfigure[Network lifetime]{
	\includegraphics[width=0.23\textwidth]{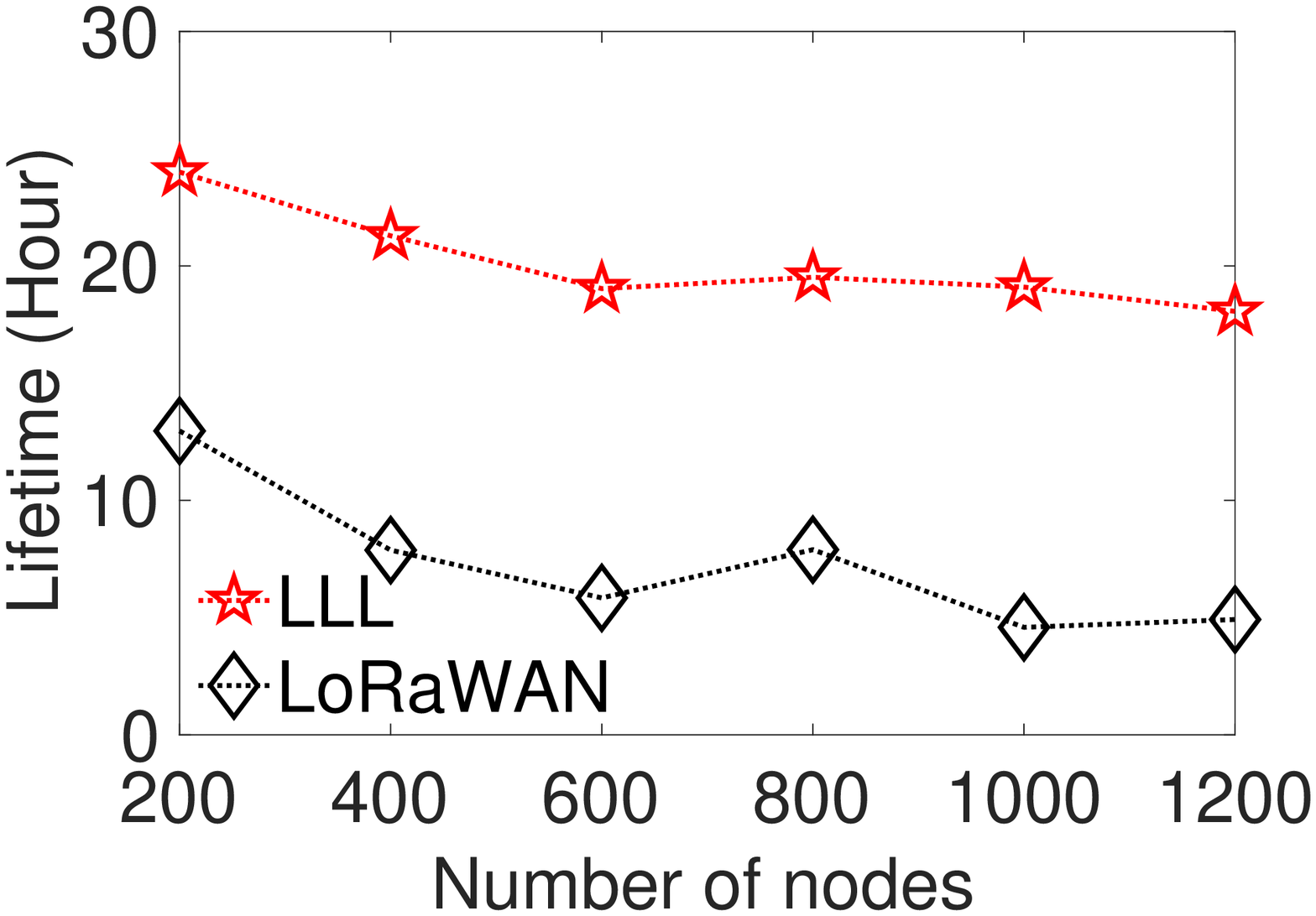} 
    \label{fig:dutycycle_lifetime}
    }%\vfill \vspace{-0.1in}
    \subfigure[Throughput]{
		\includegraphics[width=0.23\textwidth]{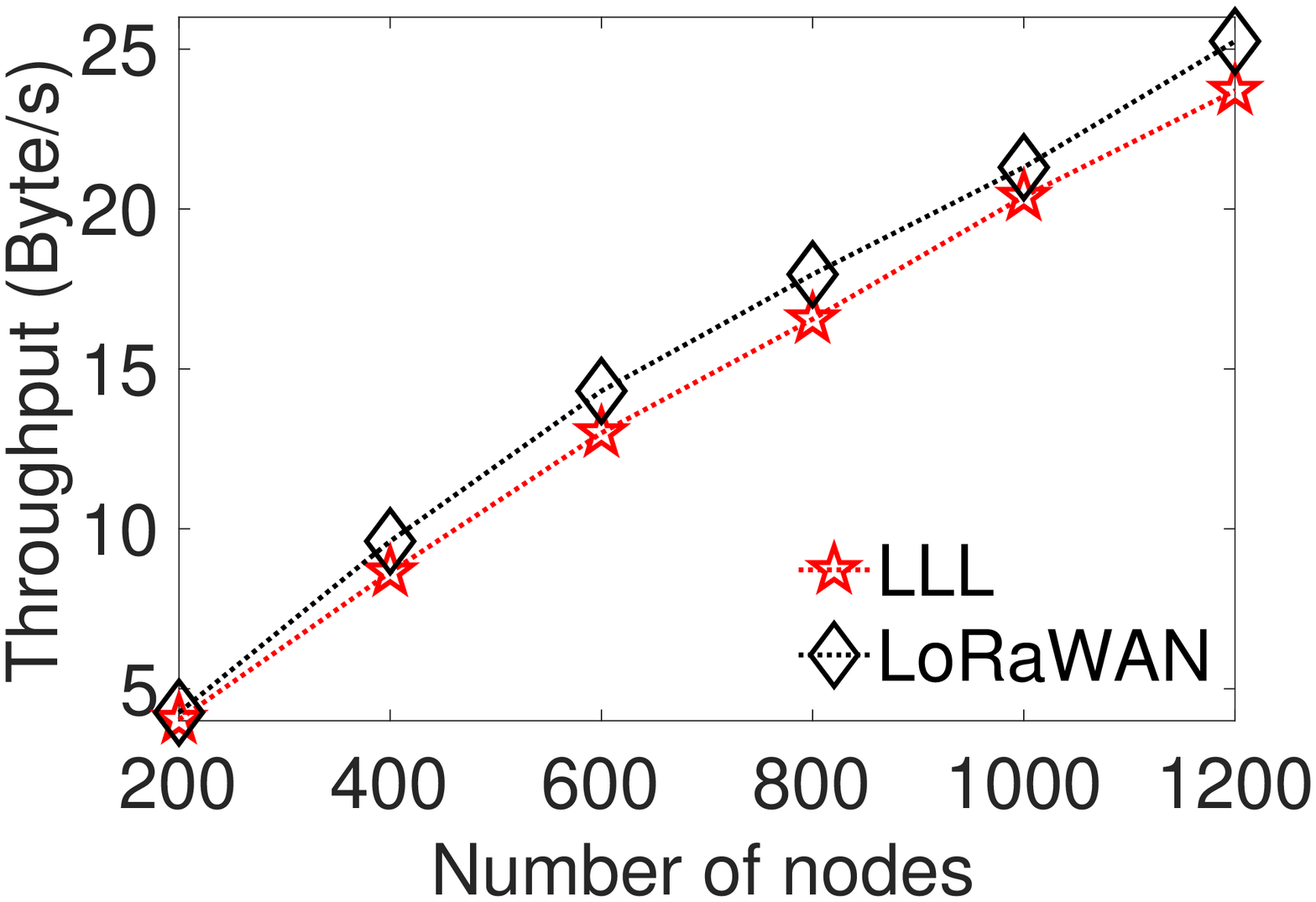}
         \label{fig:dutycycle_throughput}
	}%\quad
	\vspace{-0.1in}
	 \caption{Lifetime and throughput under varying number of nodes in EU band}
	\label{fig:vary_nodes_dutycycle}
	\vspace{-0.15in}
\end{figure}

We have evaluated our approach in the EU band using 8 channels ranging from 868 MHz to 869.55 MHz under varying number of nodes. Duty-cycle for each channel was set according to the regulatory rules. As such, the same channel cannot be reused for multiple transmissions immediately. To comply with this, we change our channel assignment in offloading and lading mode such that after each transmission, an affluent node asks the depleting nodes through ACKs to change to the same channel for the next transmission. However, in case of more than one depleting nodes with different transmission patterns, this introduces an overhead as ACKs cannot be transmitted to all of them immediately. \revise{Thus, in Figure~\ref{fig:vary_nodes_dutycycle} we see a slight decrease in lifetime and throughput for our approach. While throughput of both approaches increases with the number of nodes in Figure~\ref{fig:dutycycle_throughput}, the throughput for our approach is slightly lower than LoRa. However, as shown in Figure~\ref{fig:dutycycle_lifetime}, we observe that our approach is significantly better than LoRaWAN in terms of lifetime. On average our approach has lifetime of 20.2 Hours compared to 7.35 Hours in LoRa.} %On average the throughput of our approach is 14.38 Byte/s, whereas that of LoRaWAN is 15.44 Byte/s. . %For 600 nodes, our approach has lifetime of 19.03 Hours, where  LoRaWAN has lifetime of 5.83 Hours.  

\subsection{Experiments}

In this section, we conduct a proof-of-concept experiment of our design with the commercially available LoRa nodes, gateway, and network server to show that the simulation results are consistent also in our real testbed, i.e., the measured network lifetime increases while maintaining a similar throughput.

%In this section, we provide the experimental evaluation of our design with the commercially available LoRa nodes, gateway, and network server. \revise{The following implementation, setup, and results, are mainly meant to conduct a proof-of-concept and feasibility experiment to show that the simulation results are consistent also in our real testbed, i.e., the nodes' measured lifetime increases but maintain a similar throughput.}

% On the other hand, as explained in the following sections, the recent COVID-19 situation and the lock down status of our city, department, and labs, have lead to some limitations in  terms of possibility to measure energy consumption and to run outdoor experiments.

\subsubsection{Implementation}
In our implementation, we use the Dragino SX1276 LoRa transceiver HAT~\cite{lorahat} on Raspberry Pi 3~\cite{rpi3} as the LoRa nodes. On the other hand, we use the RAK2245 HAT~\cite{gatewayhat} on Raspberry pi 3 to build a gateway. \edit{Specifically, each LoRa node runs a custom built version (for our protocol) of the LMIC 1.6 LoRa/LoRaWAN library~\cite{lmic}} and the Raspberry Pi hosting the gateway runs the open-source ChirpStack LoRaWAN network server~\cite{cstack} locally. 
%For simplicity, we adopt the same simulation results on LSTM to decide on the sets of nodes that run in offloading and lading mode, respectively.

\subsubsection{Experimental Setup}
\begin{wrapfigure}{l}{4.5cm}
    \centering
    \vspace{-0.1in}
    \includegraphics[width=0.23\textwidth]{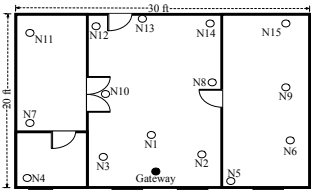} \vspace{-0.1in}
    \caption{Gateway and node's positions in indoor deployment. 
   % Note that the authors are unable to conduct wide-area outdoor experiments due to the COVID-19 lock-down situation and potential human health hazards.
    } \vspace{-0.1in}
    \label{fig:indoor}
\end{wrapfigure}

We use 15 LoRa nodes and one gateway in our experiments. Figure~\ref{fig:indoor} shows the locations of the nodes in our indoor deployment of an area of (30x20) ft$^2$, where the dark circle represents the gateway and the rest represent the nodes (labeled from N1 to N15). Note that 
we were unable to perform wide-area outdoor experiments due to the COVID-19 pandemic \cite{who}. The gateway has two half-duplex transceiver radios, and each node has one half-duplex transceiver radio. Our LoRa network operates in the 915 MHz frequency band and adopts no duty cycle requirements (as per the LoRa regional regulations~\cite{lora}). Each node in our experiments maintains a variable (i.e., storage area in the node's memory) to emulate its energy budget. At the beginning of the experiments, each node is assigned a different energy budget that ranges between 6 to 20 joules. Based on the radio states (e.g., transmit, receive, CAD, sleep, etc.), each node deducts appropriate energy from its budget \revise{ according to the energy model presented in Section~\ref{sec:lading_energy} which maintains the accuracy of the energy model given in the SX1276 datasheet~\cite{sx1276_datasheet} and that of the simulation setup in Section~\ref{sec:sim_setup}}.

In conventional LoRa mode (Class-A), each node adopts the ADR. In offloading mode, a node uses a channel bandwidth of 125 kHz, SF of 7, transmit power of 2 dBm (minimum for the SX1276 LoRa module in the LMIC library), and coding rate of $\frac{4}{5}$. Using these settings, we enable 3 to 15 nodes to transmit data in the experiments. Each node transmits its CPU temperature (2 bytes) to the gateway with the ACK bit set. In each run of the experiment (e.g., with different number of nodes), two-thirds of the nodes adopt an inter-packet interval of 2 to 4 minutes. On the other hand, one-third of the nodes send packets with an interval of 40 to 60 seconds. It may thus introduce depleting nodes in the network. Unless stated otherwise, these are our default experimental setups.

\revise{Note that the Raspberry pi 3 used as LoRa nodes in our setup has a higher static power consumption than light-weight sensors used in real deployments. We did not attach a battery to the nodes as the raspberry pi's static power usage would overwhelm the power consumed in transmissions and rapidly deplete the battery, instead the nodes were line-powered. In this proof-of-concept setup,  the supply current values from the SX1276 datasheet~\cite{sx1276_datasheet} were used to calculate the total energy consumption in our proposed Algorithms.}

%\noindent \edit{ \textbf{Rationale for Experiment Setup:} Raspberry pi 3 used as LoRa nodes in our setup has a higher static power consumption than light-weight sensors used in real deployments. We did not attach a battery to the nodes as the raspberry pi’s static power usage would overwhelm the power consumed in transmissions and rapidly deplete the battery. Furthermore, time on air for LoRa packets is in the range of milliseconds. To the best of our knowledge, the sampling rate of commercial ammeters/wattmeters do not provide the necessary accuracy to correctly measure the change in current at these small magnitudes. While the feasibility of an accurate measurement was shown very recently using some sensors like ADAfruit INA219~\cite{adafruit_paper}, due to the COVID-19 pandemic we were not able to procure such sensors before the submission. Both our campus and city were under a long-term lockdown and we were able to perform only a small-scale feasibility experiment under restrictions. Thus, we were unable to provide hardware measurements for the energy consumption. However, throughput and time on air were measured from the real hardware and supply current values from the SX1276 datasheet~\cite{sx1276_datasheet} were used to calculate the total energy consumption.}

\subsubsection{Results}
\begin{figure}[ht]
	\centering
	\vspace{-0.25in}
	\subfigure[Network lifetime]{
	\includegraphics[width=0.23\textwidth]{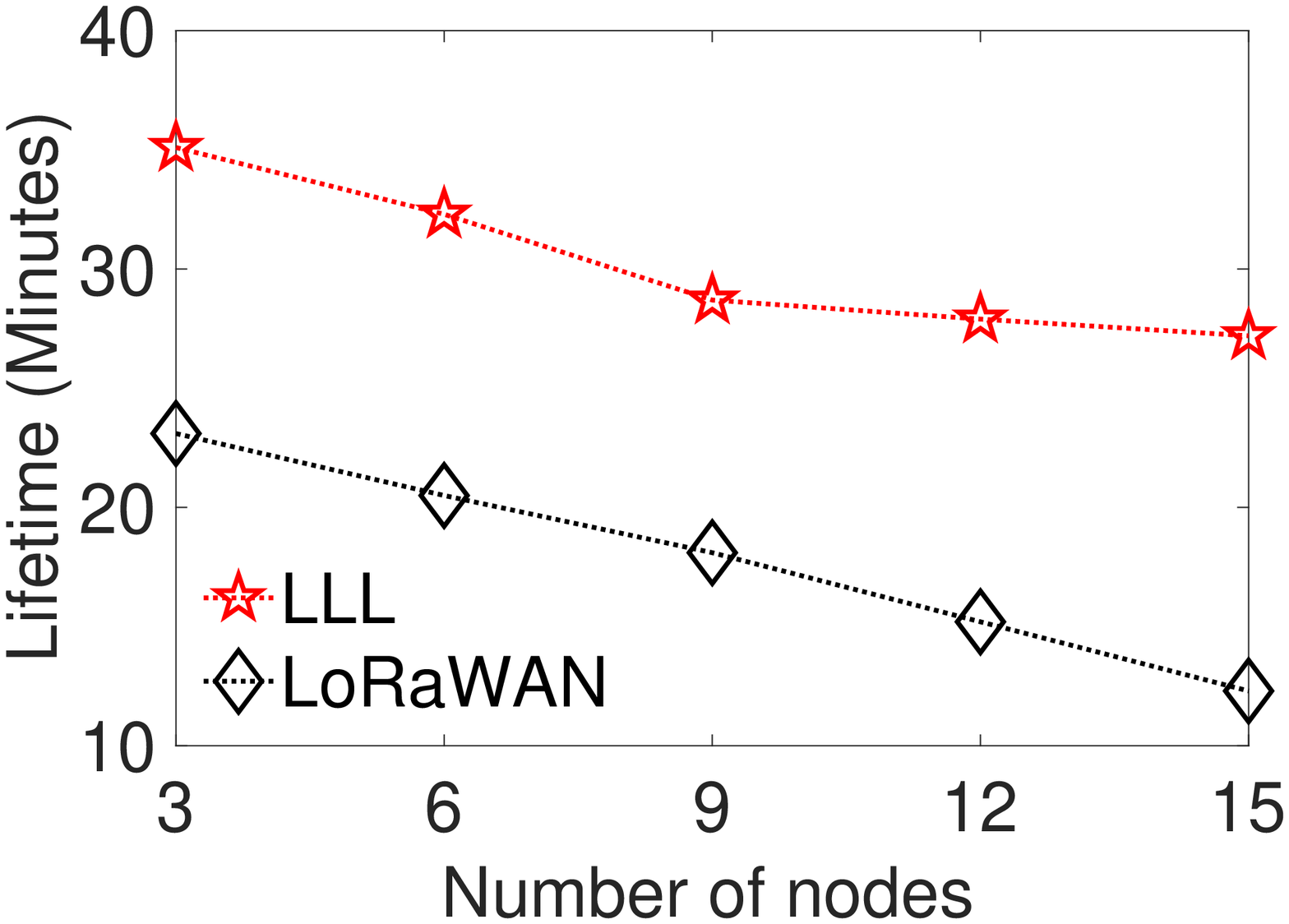}
    \label{fig:exp-lifetime}
    }%\vfill \vspace{-0.1in}
    \subfigure[Throughput]{
		\includegraphics[width=0.23\textwidth]{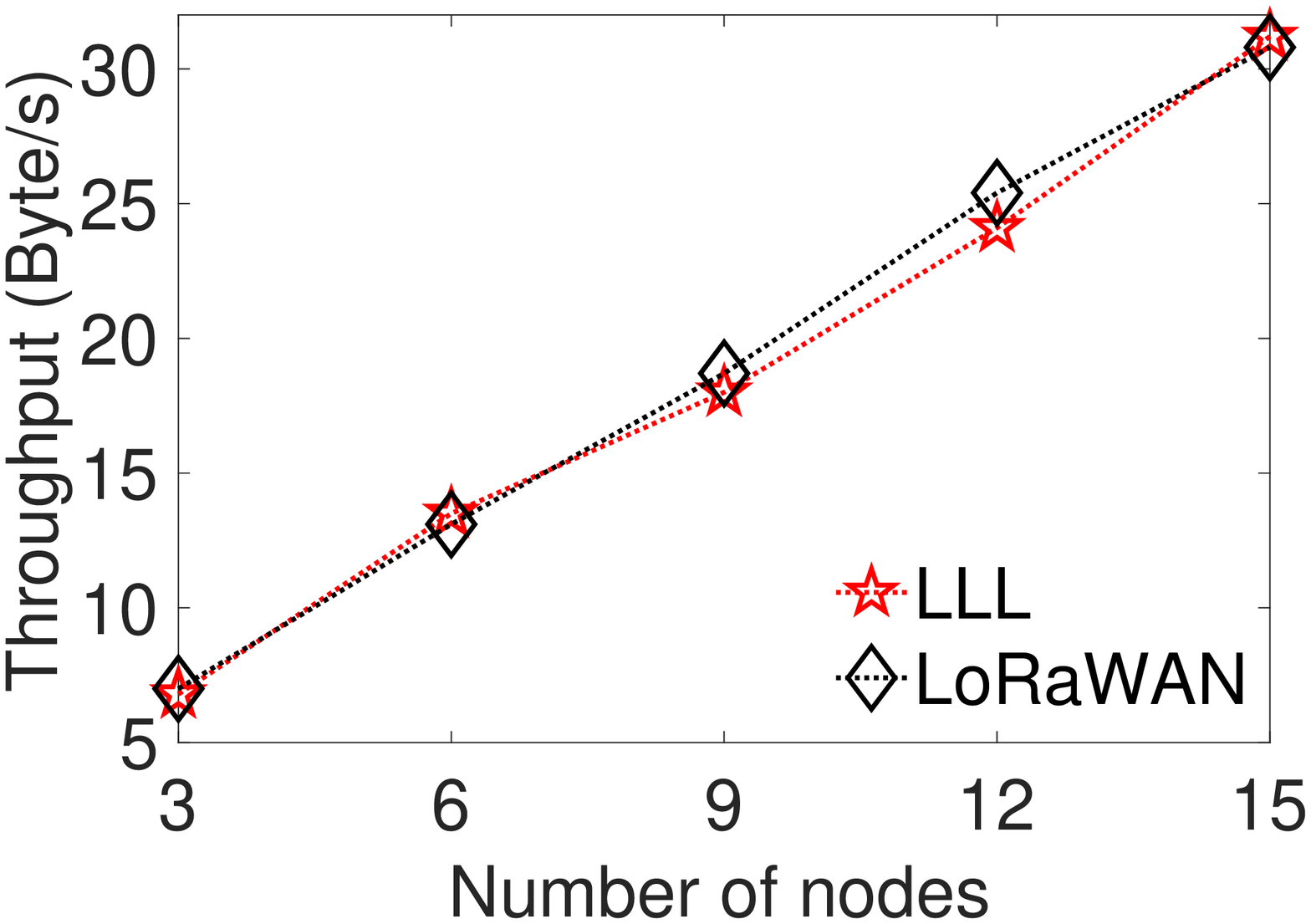}
         \label{fig:exp-tput}
	}%\quad
	\vspace{-0.1in}
	 \caption{Network lifetime and throughput under varying number of nodes}
	\label{fig:exp}
	\vspace{-0.15in}
\end{figure}

With the above settings, we evaluate the network lifetime and throughput of our design and compare with LoRaWAN.
Figure~\ref{fig:exp-lifetime} shows the network lifetime for both of the approaches with varying number of nodes. \revise{On average our approach has lifetime of 30.24 minutes, compared to 17.84 minutes in LoRa. As the nodes use a higher packet rate than simulation, the lifetime is much lower than observed in simulation.} This figure also shows that as the number of nodes increases, the lifetime in LoRaWAN decreases sharply compared to our approach.
%Specifically, when the number of nodes is 3, our approach has an overall lifetime of 35 minutes, compared to 23 minutes in LoRa. 
%. For example, for 15 nodes, our approach yields a lifetime of 27.2 minutes while it is 12.4 minutes in LoRaWAN. 
%Overall, Figure~\ref{fig:exp-lifetime} thus confirms that our approach may enable long-lasting IoT applications.
Figure~\ref{fig:exp-tput} shows the overall network throughput for both of the approaches with varying number of nodes. 
%As shown in this figure, when the number of nodes is 3, the throughput in LoRaWAN is 7 bytes/second, compared to 6.8 bytes/second in our approach. 
\revise{As shown in this figure, on average, the throughput in LoRaWAN and our approach remain competitive (18.72 vs 19 Bytes/s).}
%For example, for 15 nodes, the throughput in LoRaWAN and our approach are 30.8 and 31.2 bytes/second, respectively. 
%In general, Figure~\ref{fig:exp} shows that our approach may enable long-lasting IoT applications while providing similar network throughput as LoRaWAN. 
\revise{The results in simulations and experiments exhibit similar trends in network lifetime and throughput.}

\section{Conclusion}
%\vspace{-2mm}
\label{sec:conclusion}

%Recently LoRa networks have been used extensively to enable long-range communication between one or more gateways and numerous sensors nodes, thus enabling many IoT applications such as smart agriculture, smart city and environmental monitoring. Energy-harvesting LoRa networks have significant potential to enable seamless operation in various applications. In this paper, we propose a novel system to prolong the lifetime of an energy-harvesting LoRa network. Our approach takes advantage of extra residual energy at some nodes to assist other nodes in the network which are depleting their battery, thereby increasing the lifetime of the network. Our system achieves this using novel features which ensure low-overhead of communication and consistent packet delivery. Finally, we evaluate our approach using large scale simulations and physical experiments. Experiment results show that our approach can increase the lifetime of the LoRa network up to $4$ times, while maintaining the same throughput. 

%Recently LoRa networks have been used extensively to enable long-range communication between one or more gateways and numerous sensors nodes, thus enabling many IoT applications such as smart agriculture, smart city and environmental monitoring. Energy-harvesting LoRa networks have significant potential to enable seamless operation in various applications. 

As LoRa networks are widely being adopted for various IoT applications, prolonging their  lifetime is a very important research problem. In this paper, we have proposed  a link-layer protocol to prolong the lifetime of an energy-harvested LoRa network. Simulations and experiments show that our protocol can increase the network lifetime up to $4$ times  compared to traditional LoRa network. \revise{In our future work we intent to study how to preserve the privacy of packets while offloading.}

\section*{Acknowledgment}
This work was supported by NSF through grants CAREER-1846126 and CNS-2006467.

\bibliographystyle{IEEEtran}
\bibliography{bibfile,lorabib}

\end{document}